\documentclass{article}
\pdfoutput=1
\usepackage{jcappub}

\usepackage{multirow,longtable}
\let\subsectionautorefname\sectionautorefname

\newcommand{\Autoref}[1]{%
  \begingroup%
  \def\sectionautorefname{Section}%
  \def\subsectionautorefname{Subsection}%
  \autoref{#1}%
  \endgroup%
}

\usepackage[range-phrase=\textendash, range-units=single, separate-uncertainty=true]{siunitx}
\let\sun\odot
\DeclareSIUnit\solarmass{\ensuremath{M_\sun}}
\DeclareSIUnit\erg{erg}
\DeclareSIUnit\year{yr}
\DeclareSIUnit\parsec{pc}

\subheader{Published in JCAP}
\title{Searches for neutrino counterparts of gravitational waves from the LIGO/Virgo third observing run with KM3NeT}

\author[a]{S.~Aiello}
\author[b,bc]{A.~Albert}
\author[c]{S. Alves Garre}
\author[d]{Z.~Aly}
\author[f,e]{A. Ambrosone}
\author[g]{F.~Ameli}
\author[h]{M.~Andre}
\author[i]{E.~Androutsou}
\author[j]{M.~Anguita}
\author[k]{L.~Aphecetche}
% ...
\emailAdd{mathieu.lamoureux@uclouvain.be}
\author[w]{G.~Larosa}
\author[d]{C.~Lastoria}
\author[c]{A.~Lazo}
\author[d,1]{S.~Le~Stum}
\emailAdd{lestum@cppm.in2p3.fr}
\author[o]{G.~Lehaut}
\author[a]{E.~Leonora}
\author[c]{N.~Lessing}
\author[t,s]{G.~Levi}
\author[n]{M.~Lindsey~Clark}
\author[a]{F.~Longhitano}
\author[r]{J.~Majumdar}
\author[ad]{L.~Malerba}
\author[p]{F.~Mamedov}
\author[c]{J.~Ma\'nczak}
% ...
\emailAdd{vannoye@cppm.in2p3.fr}
\author[ba]{G.~Vasileiadis}
\author[r]{F.~Vazquez~de~Sola}
\author[n]{C.~Verilhac}
\author[g,ae]{A. Veutro}
\author[w]{S.~Viola}
\author[u,e]{D.~Vivolo}
\author[bb]{J.~Wilms}
\author[ab,r]{E.~de~Wolf}
\author[l]{H.~Yepes-Ramirez}
\author[i]{G.~Zarpapis}
\author[ad]{S.~Zavatarelli}
\author[g,ae]{A.~Zegarelli}
\author[w]{D.~Zito}
\author[c]{J.\,D.~Zornoza}
\author[c]{J.~Z{\'u}{\~n}iga}
\author[x]{N.~Zywucka}

\affiliation[a]{INFN, Sezione di Catania, Via Santa Sofia 64, Catania, 95123 Italy}
\affiliation[b]{Universit{\'e}~de~Strasbourg,~CNRS,~IPHC~UMR~7178,~F-67000~Strasbourg,~France}
\affiliation[c]{IFIC - Instituto de F{\'\i}sica Corpuscular (CSIC - Universitat de Val{\`e}ncia), c/Catedr{\'a}tico Jos{\'e} Beltr{\'a}n, 2, 46980 Paterna, Valencia, Spain}
\affiliation[d]{Aix~Marseille~Univ,~CNRS/IN2P3,~CPPM,~Marseille,~France}
\affiliation[e]{INFN, Sezione di Napoli, Complesso Universitario di Monte S. Angelo, Via Cintia ed. G, Napoli, 80126 Italy}
\affiliation[f]{Universit{\`a} di Napoli ``Federico II'', Dip. Scienze Fisiche ``E. Pancini'', Complesso Universitario di Monte S. Angelo, Via Cintia ed. G, Napoli, 80126 Italy}
\affiliation[g]{INFN, Sezione di Roma, Piazzale Aldo Moro 2, Roma, 00185 Italy}
\affiliation[h]{Universitat Polit{\`e}cnica de Catalunya, Laboratori d'Aplicacions Bioac{\'u}stiques, Centre Tecnol{\`o}gic de Vilanova i la Geltr{\'u}, Avda. Rambla Exposici{\'o}, s/n, Vilanova i la Geltr{\'u}, 08800 Spain}
\affiliation[i]{NCSR Demokritos, Institute of Nuclear and Particle Physics, Ag. Paraskevi Attikis, Athens, 15310 Greece}
\affiliation[j]{University of Granada, Dept.~of Computer Architecture and Technology/CITIC, 18071 Granada, Spain}
\affiliation[k]{Subatech, IMT Atlantique, IN2P3-CNRS, Nantes Universit{\'e}, 4 rue Alfred Kastler - La Chantrerie, Nantes, BP 20722 44307 France}
\affiliation[l]{Universitat Polit{\`e}cnica de Val{\`e}ncia, Instituto de Investigaci{\'o}n para la Gesti{\'o}n Integrada de las Zonas Costeras, C/ Paranimf, 1, Gandia, 46730 Spain}
% ...
\affiliation[n]{Universit{\'e} Paris Cit{\'e}, CNRS, Astroparticule et Cosmologie, F-75013 Paris, France}
\affiliation[o]{LPC CAEN, Normandie Univ, ENSICAEN, UNICAEN, CNRS/IN2P3, 6 boulevard Mar{\'e}chal Juin, Caen, 14050 France}
\affiliation[p]{Czech Technical University in Prague, Institute of Experimental and Applied Physics, Husova 240/5, Prague, 110 00 Czech Republic}
% ...
\affiliation[r]{Nikhef, National Institute for Subatomic Physics, PO Box 41882, Amsterdam, 1009 DB Netherlands}
\affiliation[s]{INFN, Sezione di Bologna, v.le C. Berti-Pichat, 6/2, Bologna, 40127 Italy}
\affiliation[t]{Universit{\`a} di Bologna, Dipartimento di Fisica e Astronomia, v.le C. Berti-Pichat, 6/2, Bologna, 40127 Italy}
\affiliation[u]{Universit{\`a} degli Studi della Campania ``Luigi Vanvitelli'', Dipartimento di Matematica e Fisica, viale Lincoln 5, Caserta, 81100 Italy}
% ...
\affiliation[w]{INFN, Laboratori Nazionali del Sud, Via S. Sofia 62, Catania, 95123 Italy}
\affiliation[x]{North-West University, Centre for Space Research, Private Bag X6001, Potchefstroom, 2520 South Africa}
% ...
\affiliation[ab]{University of Amsterdam, Institute of Physics/IHEF, PO Box 94216, Amsterdam, 1090 GE Netherlands}
% ...
\affiliation[ad]{INFN, Sezione di Genova, Via Dodecaneso 33, Genova, 16146 Italy}
\affiliation[ae]{Universit{\`a} La Sapienza, Dipartimento di Fisica, Piazzale Aldo Moro 2, Roma, 00185 Italy}
% ...
\affiliation[ba]{Laboratoire Univers et Particules de Montpellier, Place Eug{\`e}ne Bataillon - CC 72, Montpellier C{\'e}dex 05, 34095 France}
\affiliation[bb]{Friedrich-Alexander-Universit{\"a}t Erlangen-N{\"u}rnberg (FAU), Remeis Sternwarte, Sternwartstra{\ss}e 7, 96049 Bamberg, Germany}
\affiliation[bc]{Universit{\'e} de Haute Alsace, rue des Fr{\`e}res Lumi{\`e}re, 68093 Mulhouse Cedex, France}
% ...
\affiliation[be]{AstroCeNT, Nicolaus Copernicus Astronomical Center, Polish Academy of Sciences, Rektorska 4, Warsaw, 00-614 Poland}
\emailAdd{km3net-pc@km3net.de}

\abstract{The KM3NeT neutrino telescope is currently being deployed at two different sites in the Mediterranean Sea. First searches for astrophysical neutrinos have been performed using data taken with the partial detector configuration already in operation. The paper presents the results of two independent searches for neutrinos from compact binary mergers detected during the third observing run of the LIGO and Virgo gravitational wave interferometers. The first search looks for a global increase in the detector counting rates that could be associated with inverse beta decay events generated by MeV-scale electron anti-neutrinos. The second one focuses on upgoing track-like events mainly induced by muon (anti-)neutrinos in the GeV--TeV energy range. Both searches yield no significant excess for the sources in the gravitational wave catalogs. For each source, upper limits on the neutrino flux and on the total energy emitted in neutrinos in the respective energy ranges have been set. Stacking analyses of binary black hole mergers and neutron star-black hole mergers have also been performed to constrain the characteristic neutrino emission from these categories.}

\begin{document}

\maketitle

\section{Introduction}
\label{sec:intro}

The first detection of a gravitational wave (GW) signal from a binary compact merger~\cite{LIGOScientific:2016aoc} initiated in 2015 a new era in multi-messenger astronomy. The subsequent observation in 2017 of a GW signal from the binary neutron star merger event GW170817 and of prompt and afterglow electromagnetic emissions from the associated short gamma-ray burst~\cite{LIGOScientific:2017ync} was the first and so far unique multi-messenger observation of its kind.

Models exist of production of neutrinos from these compact mergers, especially for mergers involving neutron stars such as binary neutron star mergers (BNS)~\cite{Kimura:2018vvz} or neutron star-black hole mergers (NSBH)~\cite{Kimura:2017kan}, though some models also predict neutrino emissions from binary black hole mergers (BBH)~\cite{Kotera:2016dmp}. Although most of the studies focus on hadronic processes leading to high-energy neutrino production ($E_\nu \gtrsim \unit{\giga\electronvolt}$), thermal neutrinos in the \unit{\mega\electronvolt} regime may also be produced \cite{Kyutoku:2017wnb}.

Searches for neutrinos associated with GW signals from compact binary mergers have already been performed with other neutrino telescopes across the globe e.g., ANTARES~\cite{ANTARES:2020kxp,ANTARES:2023wcj}, IceCube~\cite{IceCube:2020xks,IceCube:2021ddq,IceCube:2022mma}, and Super-Kamiokande~\cite{Super-Kamiokande:2021dav}, without positive evidence of a common signal so far.

The KM3NeT detector, currently under construction, was taking data with a partial configuration during the third GW observation campaign in 2019-2020, allowing for a first search for neutrino counterparts. The article presents the dedicated analyses that have been developed for the search and the first results obtained with KM3NeT data, using the latest GW public catalogs as detailed below. 

Two independent analyses have been performed, each of them optimized for the detection of a prompt signal in a short time window around the GW event, and for a specific neutrino energy range. \Autoref{sec:mev} describes the search for neutrinos in the \qtyrange{5}{30}{\mega\electronvolt} range using a similar method to the one used to detect Core-Collapse Supernovae (CCSN)~\cite{KM3NeT:2021moe}, while \autoref{sec:gev} presents the search for neutrinos with energies from GeV to TeV.

The results of both searches are presented in \autoref{sec:results}. The observations are converted into constraints on the incoming neutrino flux and on the total energy radiated in neutrinos for an isotropic emission around the source, in the relevant energy ranges, assuming a quasi-thermal distribution for \unit{\mega\electronvolt} neutrinos and a single power law for GeV--TeV neutrinos. Additionally, for the latter, a stacked analysis has been performed to constrain the typical emission from BBH and NSBH objects. Results and prospects for future observations are discussed in \autoref{sec:concl}.

\subsection{The KM3NeT neutrino telescope}

The KM3NeT Collaboration is building two large-volume neutrino detectors in the depths of the Mediterranean Sea~\cite{KM3Net:2016zxf}. They rely on the detection of the Cherenkov light induced by charged particles produced in neutrino interactions, using about 200,000 three-inch photomultiplier tubes (PMTs). The PMTs are arranged in digital optical modules~\cite{KM3NeT:2022pnv} (DOMs, with 31 PMTs each), deployed along vertical lines anchored at the sea bed, with 18 DOMs per line.

The KM3NeT/ORCA detector, located near Toulon (France), will be equipped with 115 such lines, with inter-line and inter-DOM spacings that are optimized for the detection of \unit{\giga\electronvolt}-scale neutrinos and the study of atmospheric neutrino oscillations. The KM3NeT/ARCA detector is located near Capo Passero in Sicily (Italy) and will consist of two blocks of 115 lines, with larger spacings optimized for TeV--PeV astrophysical neutrinos.

\begin{figure}[hbtp]
    \centering
    \includegraphics[width=0.5\linewidth]{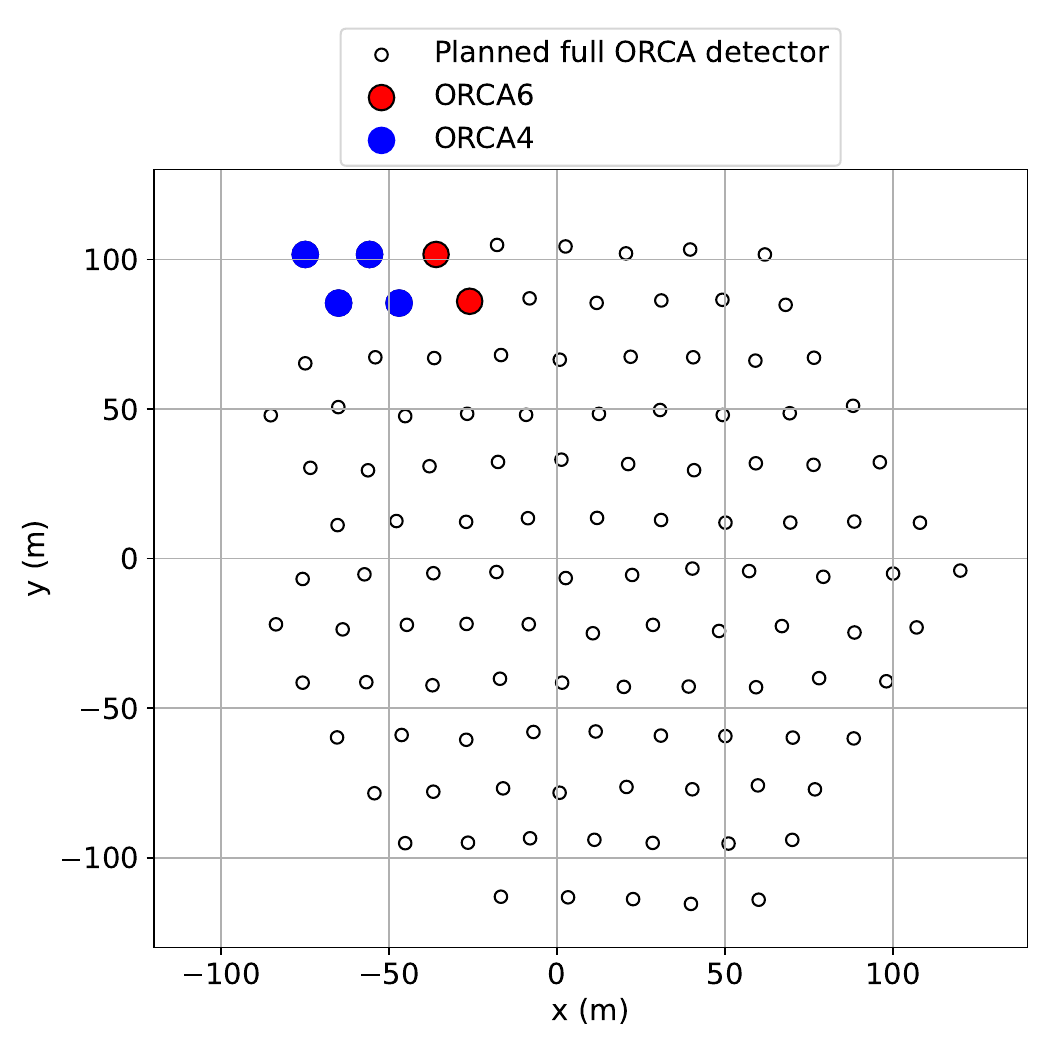}
    \caption{Footprint of the planned ORCA detector, with the ORCA4 and ORCA6 configurations highlighted in blue and red, respectively.}
    \label{fig:footprint}
\end{figure}

Detection lines are currently being deployed on both sites. At the time of the GW observations in 2019--2020, ORCA was taking data with two lines (ORCA2) before July 1, 2019, with four lines (ORCA4) during the period from July 1, 2019 to January 17, 2020, and then with six lines (ORCA6), as illustrated on the detector footprint in \autoref{fig:footprint}. The ORCA2 configuration is not considered in the following, as it is not large enough to perform a proper astrophysical search. The KM3NeT/ARCA detector has no data available for physics analysis during the considered period.

KM3NeT data is organized in consecutive runs of a few hours, and two main categories of neutrino events can be identified within the data. As it will be detailed in \autoref{sec:mev}, \unit{\mega\electronvolt} neutrinos produce individually a very faint signal such that they can only be detected through a global increase of the detector counting rate linked to many \unit{\mega\electronvolt} neutrinos interacting simultaneously. For higher energies (GeV and above), the total amount of deposited Cherenkov light distributed over multiple DOMs is sufficient to define unambiguously an event. This event would eventually be associated with an individual neutrino candidate.

\subsection{The gravitational wave catalogs}

The paper focuses on candidate binary mergers detected during the third observing run (O3) of the LIGO and Virgo GW interferometers reported in the three catalogs:
\begin{itemize}
    \item GWTC-2 \cite{LIGOScientific:2020ibl}: it reports 39 significant detections made during O3a, the first half of O3, running from April to September 2019.
    \item GWTC-2.1 \cite{LIGOScientific:2021usb}: this is an update of GWTC-2 with eight additional events detected during O3a but not reported in the previous catalog.
    \item GWTC-3 \cite{LIGOScientific:2021djp}: it reports 35 significant detections during O3b, the second half of O3, from November 2019 to March 2020. In addition, the catalog reports seven marginal candidates, out of which GW200105\_162426 has been identified as an interesting NSBH candidate and is therefore included in the analysis, making the total number of selected events 36 for GWTC-3.
\end{itemize}

The data releases provided by the LIGO-Virgo Collaboration contain detailed information for each GW event including its timing $t_{\rm GW}$, the localization skymap $\mathcal{P}(\Omega)$, and the full posterior samples with all relevant source parameters: direction $\Omega$, luminosity distance $D_L$, masses $m_{1,2}$, and total radiated energy in GWs $E_{\rm GW}$ (defined as the difference between the final object mass and the sum of the masses of the initial objects). The different categories of events (BBH, NSBH, BNS) are determined on the basis of the individual masses of the merging objects, with a chosen boundary at $m=\qty{3}{\solarmass}$ separating between neutron stars (below) and black holes (above). Other parameters are used in the follow-up analyses detailed in the following sections.

The ORCA4 data-taking period overlaps with 19, 6, and 17 GW events in the GWTC-2, GWTC-2.1, and GWTC-3 catalogs, respectively, while ORCA6 overlaps with 19 GW events reported in the GWTC-3 catalog, for a total of 61 GW events. The remaining 20 (2) events in the GWTC-2 (GWTC-2.1) catalogs occurred before ORCA4 started.

\section{Search for neutrinos in the 5--30 MeV energy range}
\label{sec:mev}

In a DOM, a hit is recorded when the voltage of a PMT rises above a 0.3 photoelectron threshold. Every hit is recorded and digitized before being grouped in segments of \qty{100}{ms} called timeslices. Most of the recorded hits originate from optical noise due to radioactive decays in seawater, mainly ${}^{40}$K (around $\qty{7}{\kilo\hertz}$ per PMT), bioluminescence which can cause localized increases up to the \unit{\mega\hertz} range, and atmospheric muons, as characterized in \cite{KM3NeT:2014wix,KM3NeT:2015qpx}.

In the \qtyrange{5}{30}{\mega\electronvolt} energy range, KM3NeT is mainly sensitive to the inverse beta decay channel, where electron anti-neutrinos interact with free protons in the water to produce low-energy positrons. Those secondary particles emit Cherenkov light for only a few tens of centimeters. As the distance between optical modules is optimized for the detection of higher energy neutrinos (above few \unit{\giga\electronvolt}), one such neutrino would only produce hits in a single DOM. Optical noise would also produce such a localized signal, making it indistinguishable from a single neutrino interaction.

Therefore, MeV neutrinos can only be detected as a global increase in the rate of coincidences between PMTs in single DOMs. The current method implemented to detect MeV neutrinos with the KM3NeT detector is optimized for the detection of a Galactic or near-Galactic CCSN, as described in \cite{KM3NeT:2021moe,KM3NeT:2021oaa}. The method assumes a quasi-thermal neutrino distribution and an emission duration of around $\qty{500}{\milli\second}$, similar to what is expected for CCSN.

To reduce the contamination from optical noise, the concept of coincidence is defined. A coincidence consists of at least four hits within one DOM and with PMTs within a 90-degree opening angle, with all the hits in a time window of \qty{10}{\nano\second}. The coincidence level is then defined as the number of coincidences over the whole detector in a sliding window of 5 timeslices (with a total duration of \qty{500}{\milli\second}) and is estimated every \qty{100}{\milli\second}. This parameter is expected to follow a Poisson distribution, characterized by a parameter $\bar{b}_c$ referred to as the ``expected background'' in the following.

The search focuses on prompt neutrino emission coincident with the GW event, with similar timing as expected for a CCSN \cite{Hansen:2019giq}. Existing models for prompt MeV neutrino emission from binary mergers have most of the signal in tens of milliseconds after the merger~\cite{Sekiguchi:2011zd,Kyutoku:2017wnb}, though the signal may extend up to a few seconds. However, to determine the time window during which the temporal correlation search is performed, it is necessary to consider the time-of-flight difference $\Delta T_{\rm flight}$ between gravitational waves and \unit{\mega\electronvolt} neutrinos (assuming the former travel at the speed of light):
\begin{equation}
  \Delta T_{\rm flight} < D_{\rm max}\left( \frac{1}{v_\nu} - \frac{1}{c} \right) = \frac{D_{\rm max}}{c} \left( \sqrt{1 + \frac{m_\nu^2 c^4}{p_\nu^2 c^2}} - 1 \right) \approx \frac{1}{2} \frac{D_{\rm max}}{c} \frac{m_\nu^2 c^4}{E_\nu^2},
\end{equation}
where $D_\text{max}$ is the estimated distance of the farthest GW source, $v_\nu$, $m_\nu$, $p_\nu$, and $E_\nu$ are respectively the speed, mass, momentum, and energy of the neutrino. Given the current constraints on the neutrino mass~\cite{Palanque-Delabrouille_2015,Capozzi:2021fjo} and the distances of GW events reported in considered catalogs, it is found that $\Delta T_{\rm flight} < \qty{2}{\second}$. 

The search window should be as short as possible to keep the trial factor (number of times the coincidence level is computed) low. A fixed time window of \qty{2}{\second} after the GW event, covering solely the bulk of the expected prompt signal and the maximum expected time-of-flight difference, is thus considered in the following.

The search consists of three steps: the selection of runs with sufficient quality, the characterization of the background, and the search for a time-correlated signal in the \qty{2}{\second} window.

\subsection{Run selection}

The characterization of the coincidence levels due to the expected background is needed to perform the analysis. For each GW event, all data from the run covering the GW time is used, in addition to the specific coincidence levels during the corresponding \qty{2}{\second} time window. For five of those GW events, data acquisition issues prevented data from being retrieved. In order to remove occasional anomalies such as sparking PMTs, which may result in multiple coincidences happening in a single DOM during \qty{100}{\milli\second}, a quality score is computed in association with every coincidence level. The quality score, as described in \cite{KM3NeT:2021oaa}, checks the consistency between the number of coincidences and the number of DOMs detecting at least one coincidence. A low score would indicate that one or several DOMs are producing an anomalous number of coincidences, which is not compatible with the expected background or signal. One additional GW event was removed from the studied sample due to a low-quality score within the \qty{2}{\second} time window, taking the total number of disregarded GW events to six. The analysis described below focuses on the 55 remaining GW events.

\subsection{Background characterization}

\begin{figure}[hbtp]
    \centering
    \includegraphics[width=0.7\linewidth]{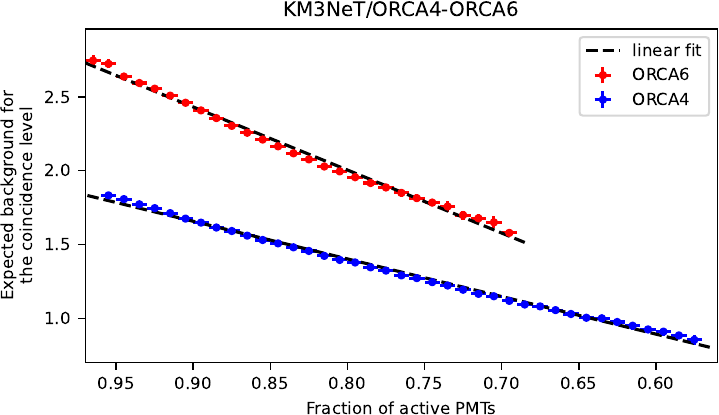}
    \caption{expected background of the coincidence level as a function of the fraction of active PMTs for ORCA4 (blue) and ORCA6 (red). The crosses indicate averaged values over the full periods, and the dashed lines are linear fits to these points.}
    \label{fig:abe}
\end{figure}

In the sea bioluminescence may lead to a localized increase of the hit rates up to the \unit{\mega\hertz} level, causing the need to veto PMT with rates above $\sim \qty{100}{\kilo\hertz}$ with the embedded electronics of the DOMs \cite{KM3NeT:2019shc}. This leads to a non-constant number of active PMTs over the whole detector, which also causes variation in the expected background. The typical timescale of those variations is a few hours. The relation between these quantities is shown in \autoref{fig:abe}, where every dot is the computed expected background averaged for the whole detector, for a given range of fraction of active PMTs, as obtained from $\sim 200$ runs, uniformly distributed in the ORCA4 and ORCA6 data-taking periods. As expected, a smaller fraction of active PMTs leads to a smaller expected background.

For each run containing a GW event, the expected background is inferred from the observed fraction of active PMTs based on a linear fit as shown in \autoref{fig:abe}. The agreement between this expectation and the observed rate has been found to be sufficient for most of the runs containing a GW event, except the six ones between December 19, 2019 and January 25, 2020. The disagreement is due to a network issue between the ORCA detector and the shore station. Instead of using the linear fit, the expected background is directly taken from data for the six runs in question. As the fraction of active PMTs is relatively stable in the runs of interest, this expected background estimation is adequate.

\subsection{Statistical analysis}

\begin{figure}[hbtp]
    \centering
    \includegraphics[width=0.48\textwidth]{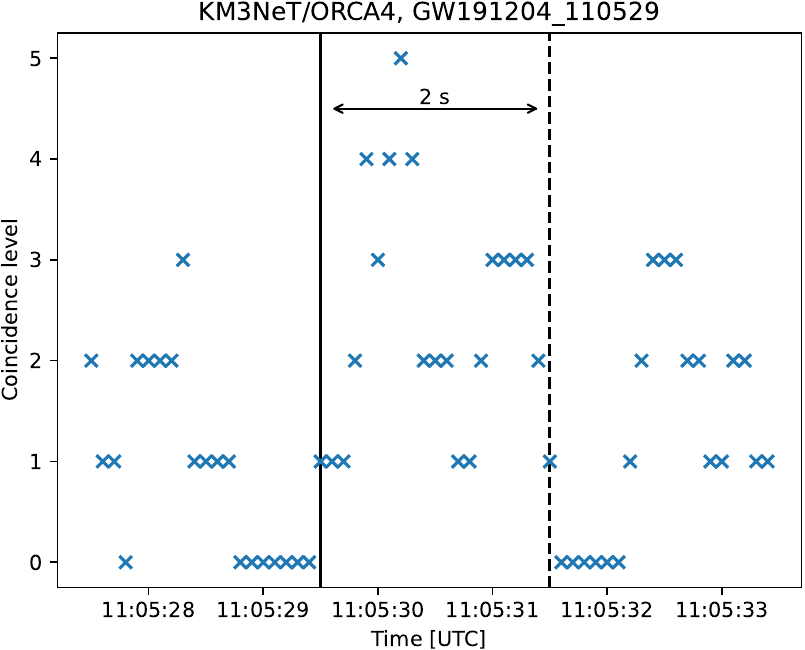}\hspace{0.03\textwidth}
    \includegraphics[width=0.48\textwidth]{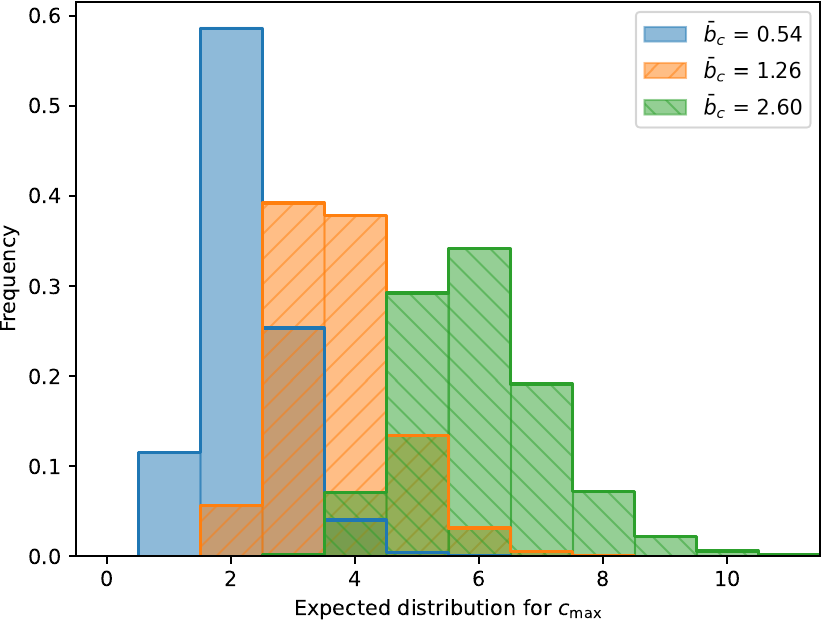}
    \caption{Timeline of the coincidence levels around GW101204\_110529 (left) and distribution of the maximum coincidence level $c_{\max}$ for different values of the expected background $\bar{b}_c$ (right). On the left, the solid black line indicates the GW event time, and the dashed black line is the end of the \qty{2}{\second} time window during which the search is made.}
    \label{fig:clevel}
\end{figure}

\begin{figure}[hbtp]
    \centering
    \includegraphics[width=0.49\textwidth]{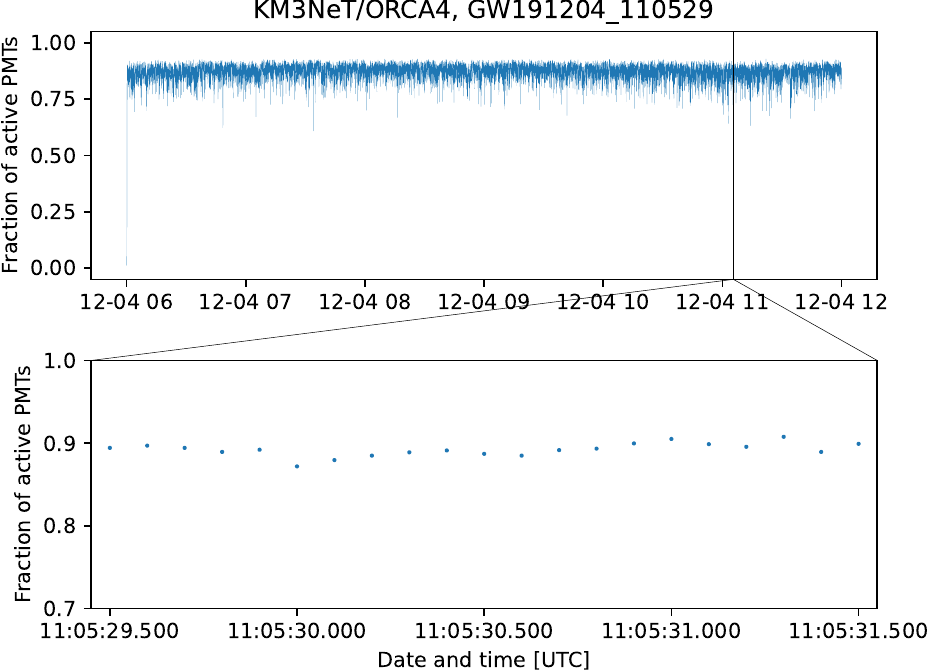}
    \includegraphics[width=0.49\textwidth]{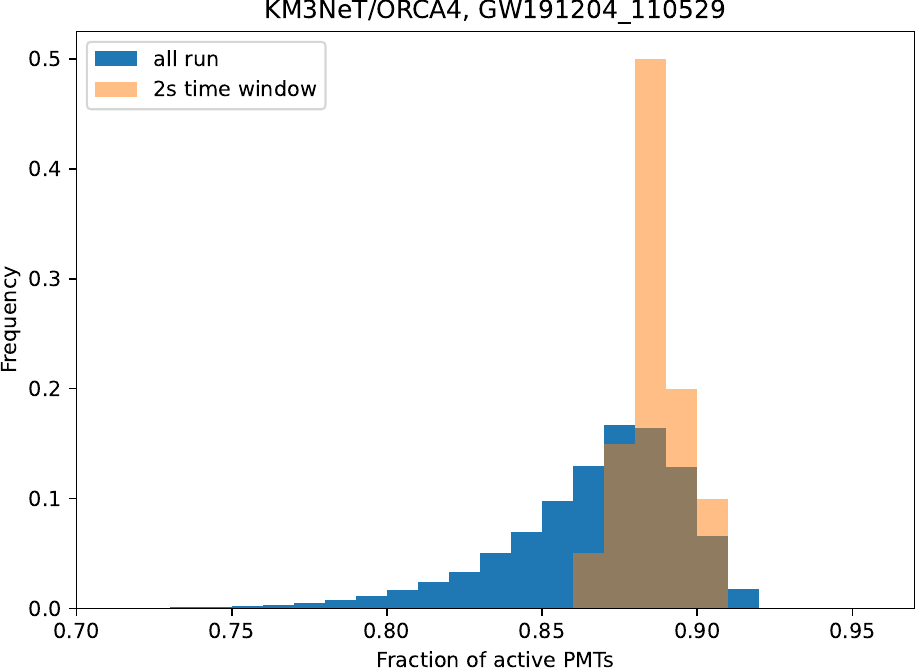}
    \caption{Time series (left) and distribution (right) of the fraction of active PMTs for the run covering GW191204\_110529. On the left, the top plot shows the variability of the fraction of active PMTs during the run while the bottom plot is a zoom on the \qty{2}{\second} time window starting from the GW event time. On the right, the distribution of the fraction of active PMTs is shown in blue for every timeslice of the run and in orange for the 20 timeslices inside the \qty{2}{\second} time window.}
    \label{fig:frac_apmts}
\end{figure}

As there is no event-by-event direction reconstruction of neutrinos at the MeV scale, the analysis consists only of a time coincidence search. For every GW event the 20 coincidence levels in the $\left[t_{\rm GW}, t_{\rm GW}+\qty{2}{\second}\right]$ time window (every \qty{100}{\milli\second} in the search window) are retrieved, as shown on the left panel of \autoref{fig:clevel}, and the maximum coincidence level $c_{\max}$ is extracted. Pseudo-experiments are then generated using the expected background $\bar{b}_c$ inferred from the observed fraction of active PMTs (\autoref{fig:frac_apmts}) averaged over the \qty{2}{\second} time interval. From those pseudo-experiments, the expected distributions of $c_{\max}$ are computed assuming only background and background + signal. Example distributions of $c_{\max}$ for different measured values of $\bar{b}_c$ are shown on the right panel of \autoref{fig:clevel}. Comparing these distributions with the observed $c_{\max}$ allows for the estimation of the corresponding p-value $p$ and the 90\% confidence level upper limit on the number of coincidences due to a neutrino signal $\mu^{90\%}_{\rm sig}$ using the Feldman-Cousins \cite{Feldman:1997qc} statistical approach.

In order to translate this quantity into physical limits, the number of expected signal events $\mu_{\rm sig, full det}(E_0, D_{L,0})$ is computed for a reference CCSN at a distance $D_{L,0}$ with a neutrino fluence $\Phi_0$ and a total released neutrino energy $E_0$, in a full ORCA detector ($115 \times 18$ DOMs), and with perfect efficiency ($\eta = 1$, where $\eta$ is the ratio of the measured expected background to the one when all PMTs are active). By correcting for the number of active DOMs $n_{\rm active~DOMs}$ and for $\eta$, an upper limit is obtained on the total neutrino fluence $\Phi^{90\%}$ and on the total energy emitted in MeV neutrinos by the source $E_{\rm tot,\nu}^{{\rm iso,} 90\%}$:
\begin{align}
    \Phi^{90\%} &= \frac{115 \times 18}{n_\text{active DOMs}} \times \frac{1}{\eta} \times \frac{\mu^{90\%}_{\rm sig}}{\mu_{\rm sig, full det}(E_0, D_{L,0})} \times \Phi_0 \label{eq:limits:vle} \\
    E^{{\rm iso,} 90\%}_{\rm tot,\nu{}} &= \frac{115 \times 18}{n_\text{active DOMs}} \times \frac{1}{\eta} \times \frac{\mu^{90\%}_{\rm sig}}{\mu_{\rm sig, full det}(E_0, D_{L,0})} \times \frac{D_L^2}{D_{L,0}^2} \times E_0
\end{align}
The reference values have been computed from refined simulations based on the work done in \cite{KM3NeT:2021moe,KM3NeT:2021oaa}, assuming a quasi-thermal emission of electron anti-neutrinos: $\mu_{\rm sig, full det}(E_0, D_{L,0})=132.5$, \mbox{$D_{L,0}=\qty{10}{\kilo\parsec}$}, $\Phi_0=\qty{8.2e10}{\per\square\centi\meter}$, and $E_0=\qty{3e53}{\erg}$.

\section{Search for neutrinos in the GeV--TeV energy range}
\label{sec:gev}

The search focuses on track-like events, mostly generated by muons produced in charged-current (CC) interactions of muon (anti-)neutrinos in the vicinity of the detector. Other event topologies are not investigated in this search.

The muon direction can be reconstructed by fitting the PMT hit patterns to the expected Cherenkov emission \cite{Melis:2017f3}. Only tracks reconstructed as upgoing or close to horizontal (i.e., with a reconstructed zenith direction $\theta$ such as $\cos(\theta) > -0.1$) are selected, in order to significantly reduce the bulk of background events caused by downgoing atmospheric muons. After this selection, the remaining backgrounds affecting the search for cosmic neutrinos are atmospheric neutrinos and atmospheric muons wrongly reconstructed as upgoing. At this level, the muon contribution is still dominant as it represents more than 99\% of the observed event rate.

To further reduce the background, only events in time coincidence and in a direction compatible with the GW localization are considered. The time correlation is performed by selecting events in a time window $[t_{\rm GW} - \qty{500}{\second}, t_{\rm GW} + \qty{500}{\second}]$, a conservative estimate of the expected delay between the high-energy neutrino and the GW emission~\cite{Baret:2011tk}. This time window is much larger than the one employed in \autoref{sec:mev} as there is no problem with trial factor for the GeV--TeV search and it is therefore possible to probe not only prompt neutrino emission but also precursor or delayed processes. The source is assumed to be located within the region $\mathcal{R}_{90}$ containing 90\% of the GW probability as built directly from the GW skymap $\mathcal{P}_{\rm GW}$. Then, the space correlation criterion corresponds to considering only events reconstructed with direction $\vec{x}$ within $\mathcal{R}_{90}^{+}$ defined as:
\begin{equation}
    \mathcal{R}_{90}^{+} = \left\{ \vec{x} \, \vert \min_{\vec{d} \in \mathcal{R}_{90}}\left(\arccos{(\vec{x} \cdot \vec{d} )}\right) \leq \Delta\phi \right\}.
\end{equation}
This extension aims to cover the detector's angular resolution and corresponds approximately to the 90\% containment angle. In the following, $\Delta\phi$ is fixed to $\qty{30}{\degree}$; such a large value is due to the small size of ORCA4 and ORCA6 detectors which leads to a large tail in the angular error distribution, as illustrated in \autoref{fig:selec_perfs} in \autoref{sec:gev:evtsel}. It should be significantly reduced with the expansion of the detector.

The analysis pipeline consists of three steps: a pre-selection of data to be analyzed according to its quality (\autoref{sec:gev:runsel}), an optimized event selection (\autoref{sec:gev:evtsel}), and a statistical analysis to extract observation significance or upper limits on the neutrino emission (\autoref{sec:gev:stat}).

\subsection{Run selection}
\label{sec:gev:runsel}

Careful checks have been implemented to ensure data quality and data-taking stability around each GW event. Conservative cuts are applied to remove all runs with non-stable trigger rates, or with other issues in terms of data quality, acquisition, or calibration. It reduces the considered total livetime (entire period of data taking, also beyond O3) from 181 to 174 days for ORCA4 and from 366 to 343 days for ORCA6.

It excludes nine GW candidates for which a follow-up is not possible as the corresponding detector runs are not selected. Furthermore, two additional GW events (GW200224\_222234 and GW200311\_115853) are excluded as they have been constrained by GW observations as being fully above the KM3NeT horizon. A total of 50 GW sources remain, including 44 BBHs and 6 NSBHs. The number differs from the one reported for MeV neutrinos in \autoref{sec:mev}, the chosen quality criteria being different as analyses rely on separate data streams with distinct responses to data-taking conditions.

\begin{figure}[hbtp]
    \centering 
    \includegraphics[width=\linewidth]{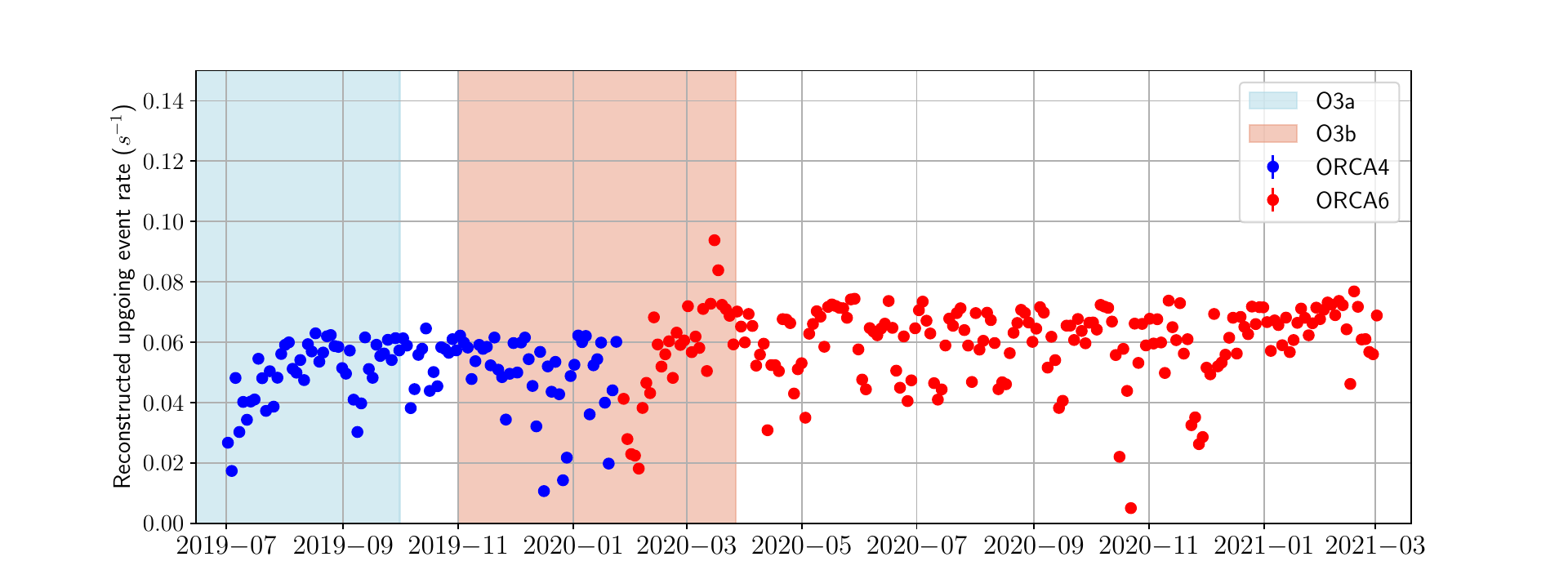}
    \caption{Rate of reconstructed upgoing events averaged over intervals of two days, for the two detector configurations ORCA4 (blue points) and ORCA6 (red points) in the data set. The shaded regions indicate the O3a and O3b periods.}
    \label{fig:reco_rates}
\end{figure}

The average rate of neutrino candidate events in the upgoing and horizontal region, in 2-day intervals, is shown in \autoref{fig:reco_rates} for the two detector configurations superimposed on the time periods covered by the GW catalogs. The main cause of fluctuations in the rate of reconstructed events is the variability of the bioluminescence at the detector. This affects the number of active PMTs as discussed in \autoref{sec:mev} which leads to fluctuations in the number of events and changes in the efficiency of track fitting.

\subsection{Analysis pipeline}
\label{sec:gev:evtsel}

The number of events in the ON-zone region in the search time window and in the direction of the GW event is compared to the expected background from mis-reconstructed atmospheric muons and atmospheric neutrinos, as estimated from OFF-zone data.

\begin{figure}[hbtp]
    \centering
    \includegraphics[width = 0.49\linewidth]{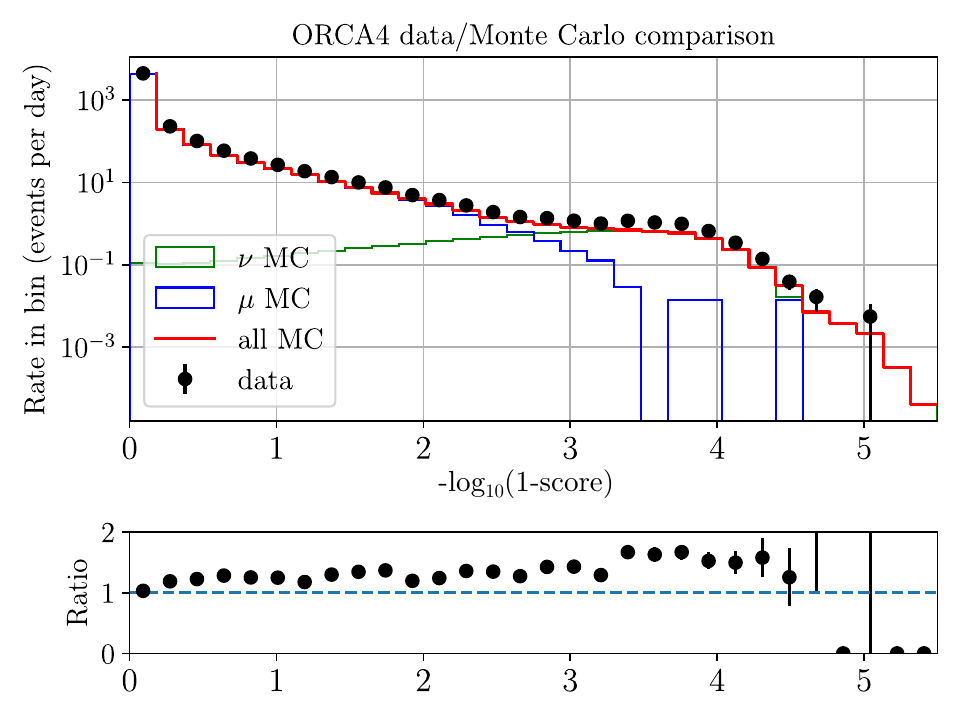}
    \includegraphics[width = 0.49\linewidth]{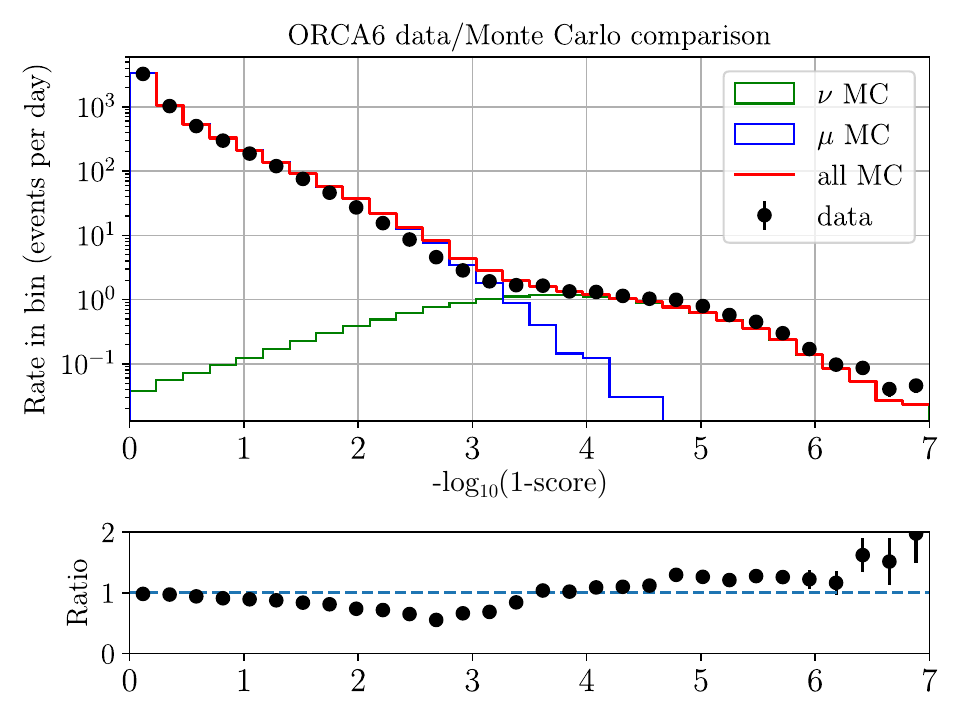}
    \caption{Comparisons of the score for upgoing events in data and Monte Carlo simulations, for the ORCA4 (left) and ORCA6 (right) detector configurations.}
    \label{fig:dataMC}
\end{figure}

A Boosted Decision Tree (BDT, based on gradient boosting~\cite{Chen:2016btl}) model is applied to select signal-like events from the dominant atmospheric muon background \cite{Assal:2021xpr}. It is trained with Monte Carlo simulations of $\nu_\mu$ CC interactions (with neutrino energies up to \qty{5}{\tera\electronvolt}) generated with gSeaGen \cite{KM3NeT:2020tvi} and muons simulated with MUPAGE~\cite{Carminati:2009fj}. The training uses a collection of 24 (14) features for ORCA4 (ORCA6), including low-level variables on the detected light as well as higher-level variables from track maximum likelihood fit results. The distribution of the final BDT scores is shown in \autoref{fig:dataMC} for data and for Monte Carlo simulations.

The ON-zone region refers to events within a $\pm \qty{500}{\second}$ time window centered on the GW event time, reconstructed as upgoing or horizontal tracks and in $\mathcal{R}_{90}^{+}$. The OFF-zone events are track-like events reconstructed within the same region in local coordinates, but at times incompatible with the GW. The OFF-zone background sample consists of a subset of runs during the same data period (ORCA4 or ORCA6) and with similar data-taking conditions with the run containing the GW event, as evaluated based on the event rate $R_{\rm loose}$ after a loose cut on the BDT score. Runs with rates in the range $[R_{\rm loose}^\star - \delta, R_{\rm loose}^\star + \delta]$ (where $R_{\rm loose}^\star$ is the event rate for the run containing the GW event) are selected, and the value of $\delta$ is optimized for each GW event to ensure $<10\%$ statistical uncertainties while having a representative background estimate. The remaining data of the run containing the GW time, outside the $\pm \qty{500}{\second}$ time window, is also part of the background sample. The ratio between the livetimes of the ON-zone and the OFF-zone regions is denoted $\alpha_{\rm{ON/OFF}}$.

A model rejection factor (MRF, \cite{Hill:2002nv}) minimization is used to optimize the cut on the BDT score, with the signal being defined as an all-flavor $E^{-2}$ neutrino spectrum, and the background being estimated from the OFF-zone region scaled by $\alpha_{\rm{ON/OFF}}$. The final cut may vary for each GW event so that the final expected background in the ON region depends on the detector conditions at the time of each GW. The detector effective area and acceptance after all cuts are estimated with the same $E^{-2}$ signal Monte Carlo simulations. The sky is divided into pixels using the HEALPix method~\cite{Gorski:2004by} and the direction-dependent acceptance $A(\Omega)$ is obtained for all pixels within the region $\mathcal{R}_{90}$.

\begin{figure}[hbtp]
    \centering
    \includegraphics[width = 0.49\linewidth]{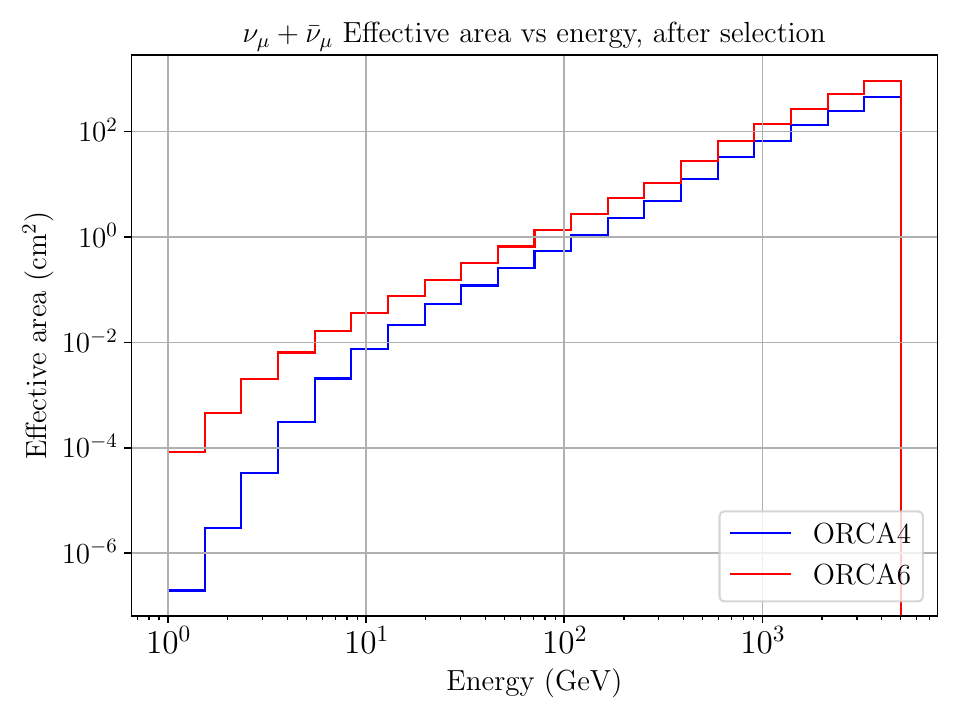}
    \includegraphics[width = 0.49\linewidth]{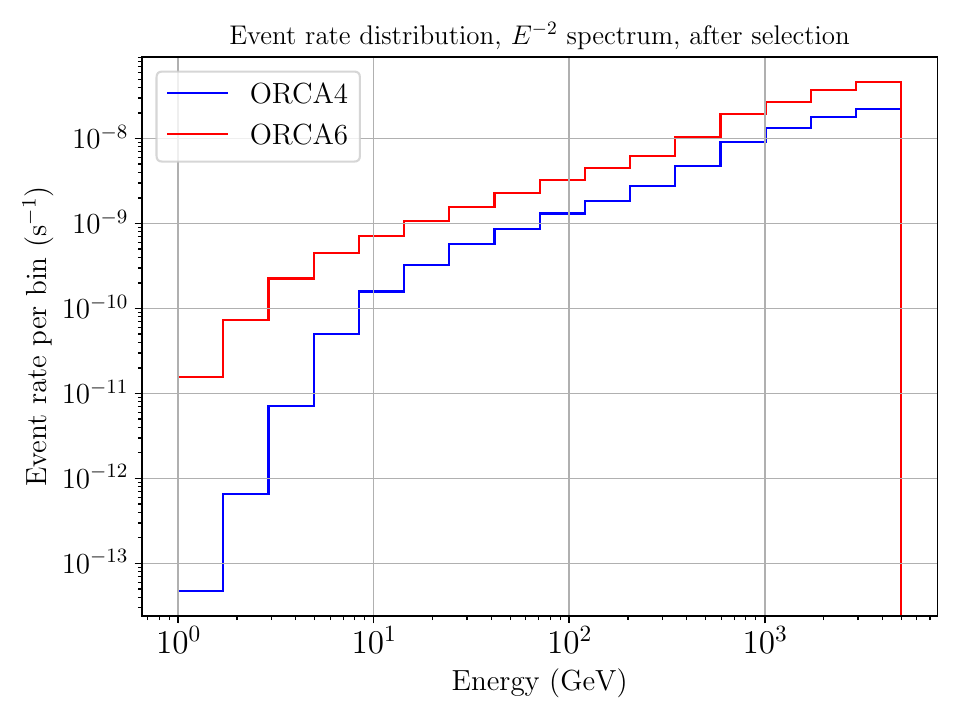}
    \includegraphics[width = 0.49\linewidth]{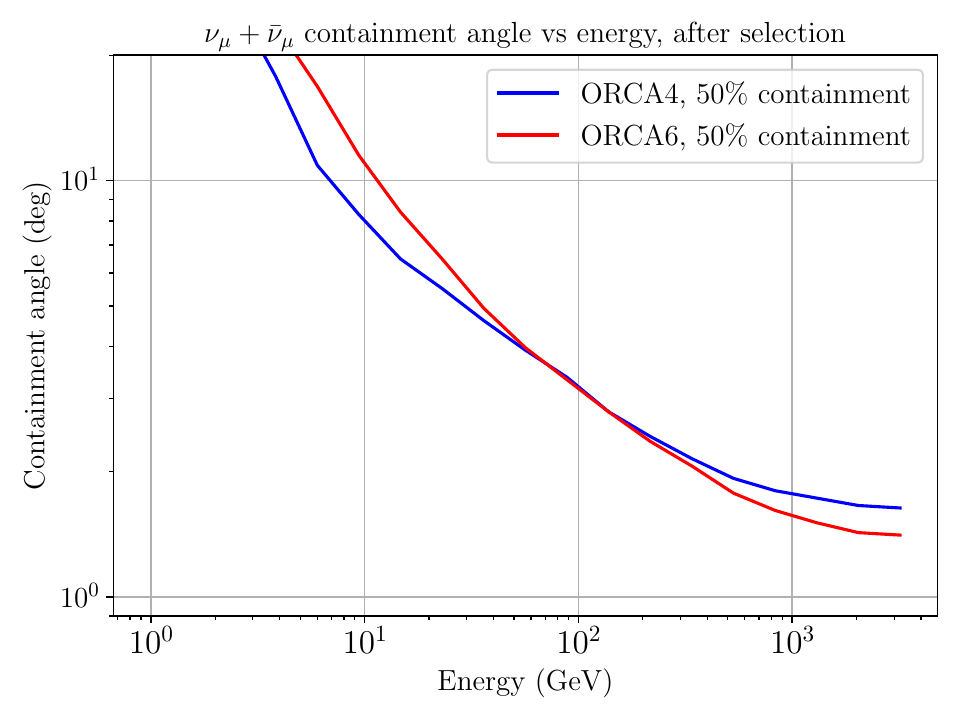}
    \includegraphics[width = 0.49\linewidth]{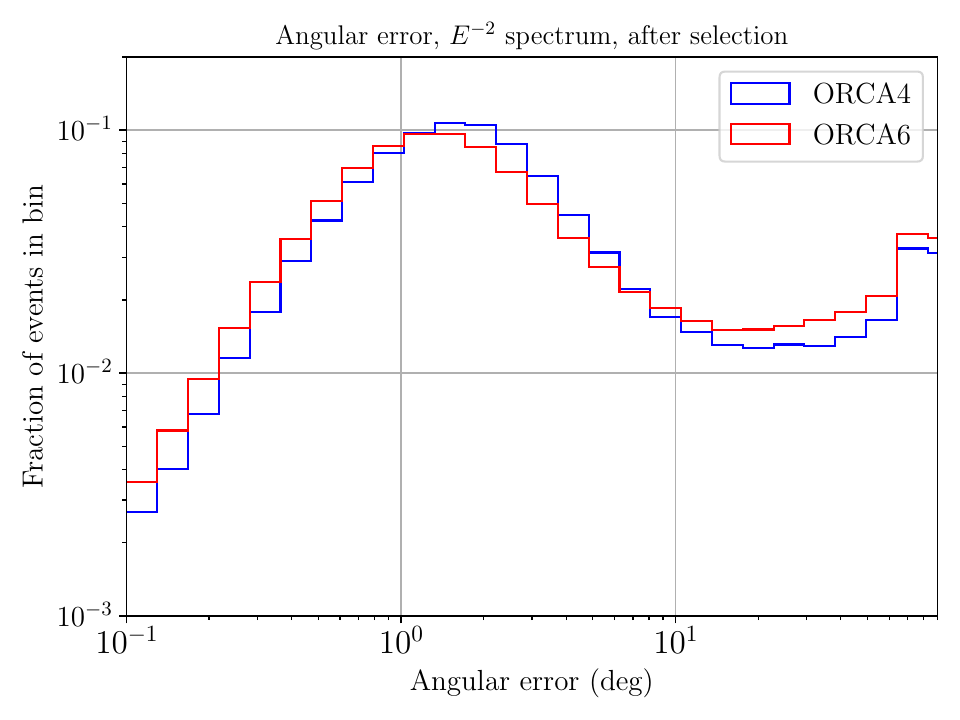}
    \caption{Performance of the $\nu_\mu + \bar\nu_\mu$ search, after optimized BDT score selection, in terms of effective area (top left), energy distribution (top right), median angular resolution (defined as the 50\% containment angle, bottom left) and angular error (bottom right). These quantities are averaged over the set of runs used in the search and shown for the two detector configurations. The two plots on the right assume an incoming neutrino flux $\num{e-4} E^{-2} \;\unit{\per\giga\electronvolt\per\square\centi\meter\per\second}$.}
    \label{fig:selec_perfs}
\end{figure}

The average effective areas, event distributions, median angular resolution (defined as the 50\% containment angle), and angular error are shown in \autoref{fig:selec_perfs} after score selection optimizations for a $\nu_\mu + \bar\nu_\mu$ flux of $\num{e-4} E^{-2}\;\unit{\per\giga\electronvolt\per\square\centi\meter\per\second}$. It should be noted that in terms of angular resolution, ORCA4 seems to outperform ORCA6 at energies below $\qty{100}{\giga\electronvolt}$ as the optimized selection is stricter in this energy range for the 4-line configuration due to its smaller size so that only higher-quality events remain. It is reflected in the event distributions, as the rate of selected low-energy events is lower. When averaged over an $E^{-2}$ spectrum, the median angular resolution for ORCA4 and ORCA6 are $\qty{1.85}{\degree}$ and $\qty{1.63}{\degree}$, respectively. It corresponds roughly to containment angles in the energy region above \qty{100}{\giga\electronvolt} on the bottom left plot of \autoref{fig:selec_perfs}, as the events at these energies are those contributing the most to the overall expected flux.

The numbers of ON-zone events $N_{\rm{ON}}$ and OFF-zone events $N_{\rm{OFF}}$, are respectively the number of events in the ON-zone and OFF-zone regions with a BDT score above the optimized cut. The mean expected number of background events in the ON-zone region is then
\begin{equation}
\bar{b} = \alpha_{\rm{ON/OFF}} \times N_{\rm{OFF}}.
\end{equation}
The corresponding Poisson p-value $p$, the Poisson probability of observing at least $N_{\rm{ON}}$ events with an expected background of $\bar{b}$ events (neglecting the related statistical uncertainty for this computation), can thus be estimated.

\subsection{Upper limit computation}
\label{sec:gev:stat}

\subsubsection*{Limits on the incoming neutrino flux for individual GW events}

The number of detected events after all cuts is compared to the background expectation from the OFF-zone region. In the absence of a significant excess, upper limits on the neutrino emission are extracted using the Bayesian framework JANG~\cite{JANG}.

Upper limits on the flux normalization $\phi$ assuming an all-flavor time-integrated neutrino spectrum ${\rm d}N/{\rm d}E = \phi \cdot (E/\unit{\giga\electronvolt})^{-2}$ are obtained, under the assumption of equipartition at Earth between neutrino flavors ($\nu_e:\nu_\mu:\nu_\tau = 1:1:1$). The corresponding likelihood is defined as:
\begin{equation}
    \mathcal{L}(N_{\rm ON} \vert b, \phi, A(\Omega)) = e^{-(b + A(\Omega) \cdot \phi)} \frac{(b + A(\Omega) \cdot \phi)^{N_{\rm ON}}}{N_{\rm ON}!},
    \label{eq:limits:lkl}
\end{equation}
where $b$ is the expected background and $A(\Omega) = a \cdot f(\Omega)$ is the direction-dependent detector acceptance, estimated with Monte Carlo simulations for each GW follow-up.

A Poisson prior $\pi(b)$ on the background with parameter $\lambda = b/\alpha_{\rm{ON/OFF}}$ encodes the information obtained from the measurements in the OFF-zone region. A 15\% (10\%) systematic uncertainty on the detector acceptance for ORCA4 (ORCA6) is reflected by defining a Gaussian prior on the acceptance normalization $a$. The GW localization skymap provided in the LIGO/Virgo catalogs is employed as prior knowledge on the source direction $\Omega$: $\pi(\Omega) = \mathcal{P}(\Omega)$. Finally, a flat prior is considered for the parameter of interest $\phi$. The posterior is then marginalized over nuisance parameters:
\begin{equation}
    P(\phi) = C \iiint_{\Omega \in \mathcal{R}_{90}^{\rm vis}} \mathcal{L}(N_{\rm ON} \vert b, \phi, A(\Omega)) \times \pi(b; N_{\rm OFF}, \alpha_{\rm{ON/OFF}}) \, \pi(a) \, \pi(\Omega) \, \pi(\phi) \, {\rm d}b \, {\rm d}a \, {\rm d}\Omega
    \label{eq:limits:posterior}
\end{equation}
where the integration is performed with Monte Carlo integration techniques and $C$ is a normalization constant. The marginalization over the source direction is only performed over the intersection $\mathcal{R}_{90}^{\rm vis}$ between the region $\mathcal{R}_{90}$ containing 90\% of the GW probability and the visible sky using the KM3NeT upgoing track sample. The 90\% upper limit on the time-integrated flux normalization $\phi^{90\%}$ is obtained by solving $\int_0^{\phi^{90\%}} P(\phi) {\rm d}\phi = 0.90$.

\subsubsection*{Limits on the total energy for individual GW events}

Similarly, upper limits on the total energy emitted in neutrinos $E^{\rm iso}_{\rm tot, \nu} = 4\pi D_{L}^2 \int_{E_{\min}}^{E_{\max}} E \times ({\rm d}N/{\rm d}E) \, {\rm d}E,$ or on the ratio between the neutrino emission and GW emission $f^{\rm iso}_\nu = E^{\rm iso}_{\rm tot, \nu}/E_{\rm GW}$, assuming an $E^{-2}$ spectrum and isotropic emission, are also derived. The procedure is similar to the ones described above with the luminosity distance $D_L$ as an additional parameter (and the total radiated energy $E_{\rm GW}$ as well for limits on $f_\nu$). The integration bounds are fixed to $E_{\min}=\qty{1}{\giga\electronvolt}$ and $E_{\max}=\qty{100}{\peta\electronvolt}$ though the obtained results may be easily scaled for different choices of bounds (e.g. $E^{\rm iso}_{\rm tot, \nu} \propto \log(E_{\max}/E_{\min})$ for an $E^{-2}$ spectrum).

\subsubsection*{Population studies}

A stacking analysis of all BBH events is also performed by combining the individual follow-up results and constraining the typical $E^{\rm iso}_{\rm tot, \nu}$ ($f^{\rm iso}_\nu$) from those objects, assuming they have the same total energy released in neutrinos (the same ratio between neutrino and GW emissions). To account for the current analysis being limited to neutrinos below the horizon (and not all-sky sensitive), stacking pseudo-experiments are performed which include each GW follow-up with a probability equal to the visibility of the corresponding $\mathcal{R}_{90}$ region. This quantity is defined as the ratio between the integrated GW probabilities in $\mathcal{R}_{90}^{\rm vis}$ and in $\mathcal{R}_{90}$:
\begin{equation}
    V = \dfrac{\int_{\mathcal{R}_{90}^{\rm vis}} \mathcal{P}(\Omega) {\rm d}\Omega}{\int_{\mathcal{R}_{90}} \mathcal{P}(\Omega) {\rm d}\Omega}. \label{eq:visibility}
\end{equation}
The quoted limit is the median value obtained from these pseudo-experiments. A similar population study is performed considering the 6 NSBH candidates in the catalogs.

\clearpage
\section{Results}
\label{sec:results}

The final results for the two analyses described in \autoref{sec:mev} and \autoref{sec:gev} are presented in \autoref{tab:results:VLE} and \autoref{tab:results:HE} respectively. No excess has been found in any of the samples and follow-ups. Therefore, only upper limits on the neutrino emission are computed and reported in the same table. For the GeV--TeV analysis, due to the low expected background rates, the computed p-values are always either 1 (if $N_{\rm ON} = 0$) or typically smaller than 0.5 (if $N_{\rm ON} > 0$) ; the values are then only provided in the latter case.

\begin{center}
\begin{longtable}[H]{|lc|S[table-format=1.2,table-auto-round]S[table-format=2]S[table-format=1.2e2,table-auto-round]S[table-format=1.2,table-auto-round]|S[table-format=1.1e2,table-auto-round]S[table-format=1.1e2,table-auto-round]|}
\caption{Summary of O3 follow-up results of the MeV analysis. For each GW event, the third and fourth columns give the expected background $\bar{b}_c$ and the maximum observed coincidence level $c_\text{max}$ during the \qty{2}{\second} window after the GW event. The next two columns report the False Alarm Rate (FAR, number of times per day one expects to observe $c_\text{max}$ coincidences originating only from background) and the p-value. The last two columns provide the obtained upper limits on the neutrino emission, in terms of the incoming fluence and the total energy emitted in neutrinos.\label{tab:results:VLE}} \\
\hline
\multirow{2}{*}{GW name} \relax & {Merger} \relax & {$\bar{b}_c$} \relax & {$c_{\max}$} \relax & {FAR} \relax & {$p$} \relax & {$\Phi^{90\%}$} \relax & {$E_{\rm tot,\nu}^{{\rm iso,} 90\%}$} \relax \\
{} \relax & {type} \relax & {} \relax & {} \relax  & {\unit{\per\day}} \relax & {} \relax & {\unit{\per\square\centi\meter}} \relax & {\unit{\erg}} \relax \\
\hline
\endfirsthead
\caption[]{(continued)} \\
\hline
\multirow{2}{*}{GW name} \relax & {Merger} \relax & {$\bar{b}_c$} \relax & {$c_{\max}$} \relax & {FAR} \relax & {$p$} \relax & {$\Phi^{90\%}$} \relax & {$E_{\rm tot,\nu}^{{\rm iso,} 90\%}$} \relax \\
{} \relax & {type} \relax & {} \relax & {} \relax  & {\unit{\per\day}} \relax & {} \relax & {\unit{\per\square\centi\meter}} \relax & {\unit{\erg}} \relax \\
\hline
\endhead
\hline
\endfoot
\multicolumn{8}{|c|}{ORCA4 period} \\
\hline
GW190701\_203306 \relax & BBH \relax & 1.44 \relax & 4 \relax & 5.04e+04 \relax & 0.7 \relax & 3.59e+10 \relax & 5.57e+61 \relax \\
GW190706\_222641 \relax & BBH \relax & 1.63 \relax & 3 \relax & 1.94e+05 \relax & 0.994 \relax & 1.67e+10 \relax & 1.19e+62 \relax \\
GW190707\_093326 \relax & BBH \relax & 1.75 \relax & 5 \relax & 2.84e+04 \relax & 0.488 \relax & 4.51e+10 \relax & 9.78e+60 \relax \\
GW190708\_232457 \relax & BBH \relax & 1.76 \relax & 5 \relax & 2.9e+04 \relax & 0.495 \relax & 4.48e+10 \relax & 1.27e+61 \relax \\
GW190719\_215514 \relax & BBH \relax & 1.7 \relax & 3 \relax & 2.1e+05 \relax & 0.996 \relax & 1.6e+10 \relax & 9.09e+61 \relax \\
GW190725\_174728 \relax & BBH \relax & 0.84 \relax & 4 \relax & 9.24e+03 \relax & 0.193 \relax & 9.39e+10 \relax & 3.79e+61 \relax \\
GW190727\_060333 \relax & BBH \relax & 0.94 \relax & 4 \relax & 1.34e+04 \relax & 0.269 \relax & 8.39e+10 \relax & 3.34e+62 \relax \\
GW190728\_064510 \relax & BBH \relax & 1.26 \relax & 3 \relax & 1.16e+05 \relax & 0.943 \relax & 2.16e+10 \relax & 5.98e+60 \relax \\
GW190731\_140936 \relax & BBH \relax & 1.66 \relax & 4 \relax & 7.54e+04 \relax & 0.839 \relax & 3.11e+10 \relax & 1.24e+62 \relax \\
GW190805\_211137 \relax & BBH \relax & 1.55 \relax & 7 \relax & 9.64e+02 \relax & 0.0219 \relax & 8.6e+10 \relax & 8.87e+62 \relax \\
GW190814 \relax & NSBH \relax & 1.66 \relax & 4 \relax & 7.54e+04 \relax & 0.839 \relax & 3.11e+10 \relax & 6.56e+59 \relax \\
GW190828\_063405 \relax & BBH \relax & 1.67 \relax & 4 \relax & 7.67e+04 \relax & 0.844 \relax & 3.1e+10 \relax & 5.14e+61 \relax \\
GW190828\_065509 \relax & BBH \relax & 1.68 \relax & 3 \relax & 2.05e+05 \relax & 0.995 \relax & 1.62e+10 \relax & 1.52e+61 \relax \\
GW190909\_114149 \relax & BBH \relax & 1.23 \relax & 1 \relax & 6.11e+05 \relax & 1.0 \relax & 1.99e+10 \relax & 1.04e+62 \relax \\
GW190917\_114630 \relax & NSBH \relax & 1.73 \relax & 4 \relax & 8.44e+04 \relax & 0.872 \relax & 2.99e+10 \relax & 5.67e+60 \relax \\
GW190924\_021846 \relax & BBH \relax & 1.77 \relax & 4 \relax & 8.98e+04 \relax & 0.889 \relax & 2.92e+10 \relax & 3.47e+60 \relax \\
GW190925\_232845 \relax & BBH \relax & 1.71 \relax & 4 \relax & 8.18e+04 \relax & 0.863 \relax & 3.02e+10 \relax & 9.56e+60 \relax \\
GW190926\_050336 \relax & BBH \relax & 1.79 \relax & 3 \relax & 2.3e+05 \relax & 0.998 \relax & 1.52e+10 \relax & 7.94e+61 \relax \\
GW190929\_012149 \relax & BBH \relax & 1.79 \relax & 3 \relax & 2.3e+05 \relax & 0.998 \relax & 1.52e+10 \relax & 2.52e+61 \relax \\
GW190930\_133541 \relax & BBH \relax & 1.72 \relax & 3 \relax & 2.14e+05 \relax & 0.997 \relax & 1.58e+10 \relax & 3.34e+60 \relax \\
GW191103\_012549 \relax & BBH \relax & 1.73 \relax & 4 \relax & 8.44e+04 \relax & 0.873 \relax & 2.99e+10 \relax & 1.07e+61 \relax \\
GW191105\_143521 \relax & BBH \relax & 1.8 \relax & 4 \relax & 9.39e+04 \relax & 0.9 \relax & 2.87e+10 \relax & 1.39e+61 \relax \\
GW191109\_010717 \relax & BBH \relax & 1.24 \relax & 4 \relax & 3.23e+04 \relax & 0.533 \relax & 4.17e+10 \relax & 2.54e+61 \relax \\
GW191113\_071753 \relax & BBH \relax & 1.75 \relax & 4 \relax & 8.71e+04 \relax & 0.88 \relax & 2.95e+10 \relax & 2.03e+61 \relax \\
GW191126\_115259 \relax & BBH \relax & 1.26 \relax & 5 \relax & 8.14e+03 \relax & 0.173 \relax & 6.26e+10 \relax & 6.01e+61 \relax \\
GW191127\_050227 \relax & BBH \relax & 1.71 \relax & 4 \relax & 8.18e+04 \relax & 0.863 \relax & 3.02e+10 \relax & 1.28e+62 \relax \\
GW191129\_134029 \relax & BBH \relax & 1.78 \relax & 1 \relax & 7.18e+05 \relax & 1.0 \relax & 1.07e+10 \relax & 2.44e+60 \relax \\
GW191204\_110529 \relax & BBH \relax & 1.64 \relax & 5 \relax & 2.24e+04 \relax & 0.409 \relax & 4.81e+10 \relax & 5.7e+61 \relax \\
GW191204\_171526 \relax & BBH \relax & 1.54 \relax & 4 \relax & 6.11e+04 \relax & 0.77 \relax & 3.36e+10 \relax & 5.19e+60 \relax \\
GW191215\_223052 \relax & BBH \relax & 1.74 \relax & 3 \relax & 2.19e+05 \relax & 0.997 \relax & 1.56e+10 \relax & 2.13e+61 \relax \\
GW191219\_163120 \relax & NSBH \relax & 0.96 \relax & 2 \relax & 2.16e+05 \relax & 0.997 \relax & 2.83e+10 \relax & 3.14e+60 \relax \\
GW191222\_033537 \relax & BBH \relax & 0.97 \relax & 2 \relax & 2.19e+05 \relax & 0.997 \relax & 2.8e+10 \relax & 9.23e+61 \relax \\
GW191230\_180458 \relax & BBH \relax & 1.05 \relax & 5 \relax & 3.88e+03 \relax & 0.0856 \relax & 1.01e+11 \relax & 6.84e+62 \relax \\
GW200105\_162426 \relax & NSBH \relax & 1.14 \relax & 4 \relax & 2.49e+04 \relax & 0.442 \relax & 6.92e+10 \relax & 1.85e+60 \relax \\
GW200112\_155838 \relax & BBH \relax & 1.15 \relax & 4 \relax & 2.56e+04 \relax & 0.451 \relax & 6.86e+10 \relax & 3.92e+61 \relax \\
GW200115\_042309 \relax & NSBH \relax & 1.15 \relax & 3 \relax & 9.49e+04 \relax & 0.902 \relax & 4.49e+10 \relax & 1.38e+60 \relax \\
\hline
\multicolumn{8}{|c|}{ORCA6 period} \\
\hline
GW200128\_022011 \relax & BBH \relax & 1.21 \relax & 3 \relax & 1.06e+05 \relax & 0.927 \relax & 6.38e+10 \relax & 2.7e+62 \relax \\
GW200129\_065458 \relax & BBH \relax & 1.23 \relax & 3 \relax & 1.1e+05 \relax & 0.934 \relax & 3.31e+10 \relax & 9.8e+60 \relax \\
GW200202\_154313 \relax & BBH \relax & 0.54 \relax & 1 \relax & 3.61e+05 \relax & 1.0 \relax & 7.53e+10 \relax & 4.63e+60 \relax \\
GW200208\_130117 \relax & BBH \relax & 1.27 \relax & 3 \relax & 1.18e+05 \relax & 0.946 \relax & 3.2e+10 \relax & 5.82e+61 \relax \\
GW200208\_222617 \relax & BBH \relax & 1.3 \relax & 4 \relax & 3.72e+04 \relax & 0.585 \relax & 5.94e+10 \relax & 3.65e+62 \relax \\
GW200209\_085452 \relax & BBH \relax & 1.29 \relax & 2 \relax & 3.19e+05 \relax & 1.0 \relax & 3.15e+10 \relax & 1.33e+62 \relax \\
GW200210\_092254 \relax & NSBH \relax & 1.25 \relax & 2 \relax & 3.07e+05 \relax & 1.0 \relax & 3.25e+10 \relax & 1.05e+61 \relax \\
GW200216\_220804 \relax & BBH \relax & 2.39 \relax & 2 \relax & 5.96e+05 \relax & 1.0 \relax & 1.53e+10 \relax & 8.09e+61 \relax \\
GW200219\_094415 \relax & BBH \relax & 1.86 \relax & 4 \relax & 1.02e+05 \relax & 0.919 \relax & 4.15e+10 \relax & 1.76e+62 \relax \\
GW200220\_061928 \relax & BBH \relax & 2.43 \relax & 4 \relax & 1.97e+05 \relax & 0.994 \relax & 1.67e+10 \relax & 2.2e+62 \relax \\
GW200220\_124850 \relax & BBH \relax & 2.58 \relax & 5 \relax & 1.03e+05 \relax & 0.922 \relax & 2.99e+10 \relax & 1.75e+62 \relax \\
GW200224\_222234 \relax & BBH \relax & 2.17 \relax & 5 \relax & 5.99e+04 \relax & 0.762 \relax & 3.56e+10 \relax & 3.81e+61 \relax \\
GW200225\_060421 \relax & BBH \relax & 2.18 \relax & 6 \relax & 2.07e+04 \relax & 0.384 \relax & 5.41e+10 \relax & 2.62e+61 \relax \\
GW200302\_015811 \relax & BBH \relax & 2.57 \relax & 3 \relax & 4.1e+05 \relax & 1.0 \relax & 1.58e+10 \relax & 1.27e+61 \relax \\
GW200306\_093714 \relax & BBH \relax & 2.45 \relax & 6 \relax & 3.35e+04 \relax & 0.546 \relax & 4.81e+10 \relax & 7.76e+61 \relax \\
GW200308\_173609 \relax & BBH \relax & 2.56 \relax & 3 \relax & 4.07e+05 \relax & 1.0 \relax & 1.43e+10 \relax & 1.52e+62 \relax \\
GW200311\_115853 \relax & BBH \relax & 2.39 \relax & 4 \relax & 1.89e+05 \relax & 0.993 \relax & 1.7e+10 \relax & 8.52e+60 \relax \\
GW200316\_215756 \relax & BBH \relax & 2.53 \relax & 5 \relax & 9.75e+04 \relax & 0.909 \relax & 3.05e+10 \relax & 1.4e+61 \relax \\
GW200322\_091133 \relax & BBH \relax & 2.6 \relax & 6 \relax & 4.24e+04 \relax & 0.634 \relax & 4.53e+10 \relax & 2.15e+62 \relax \\
\end{longtable}
\end{center}

\begin{center}
\begin{longtable}{|lc|S[table-format=3.0,table-auto-round]|S[table-format=1.3]cc|S[table-format=1.1e2,table-auto-round]S[table-format=1.1e2,table-auto-round]S[table-format=1.1e2,table-auto-round]|}
\caption{Summary of O3 follow-up results with the high-energy analysis. The second and third columns indicate the most probable merger type given the masses in the catalog and the GW localization visibility V at ORCA at the time of the merger. The next three columns report the mean expected number of background events $\bar{b}$, the observed number of events in the ON-zone region $N_{\rm ON}$, and the corresponding Poisson p-value p (in case of non-zero observations), and the last three are the obtained upper limits on the neutrino emission, in terms of the incoming time-integrated flux, the total energy emitted in neutrinos, and the ratio between neutrino and GW emissions.\label{tab:results:HE}} \\
\hline
\multirow{2}{*}{GW name} & {Merger} \relax & {$V$} \relax & {$\bar{b}$} \relax & {$N_{\rm ON}$} \relax & {$p$} \relax & {$\phi^{90\%}$} \relax & {$E_{\rm tot,\nu}^{{\rm iso,} 90\%}$} \relax & {$f_{\nu}^{{\rm iso,} 90\%}$} \relax \\
{} & {type} \relax & {\%} \relax & {} \relax & {} \relax & {} \relax & {\unit{\per\giga\electronvolt\per\square\centi\meter}} \relax & {\unit{\erg}} \relax & {} \relax \\
\hline
\endfirsthead
\caption[]{(continued)} \\
\hline
\multirow{2}{*}{GW name} & {Merger} \relax & {$V$} \relax & {$\bar{b}$} \relax & {$N_{\rm ON}$} \relax & {$p$} \relax & {$\phi^{90\%}$} \relax & {$E_{\rm tot,\nu}^{{\rm iso,} 90\%}$} \relax & {$f_{\nu}^{{\rm iso,} 90\%}$} \relax  \\
{} & {type} \relax & {\%} \relax & {} \relax & {} \relax & {} \relax & {\unit{\per\giga\electronvolt\per\square\centi\meter}} \relax & {\unit{\erg}} \relax & {} \relax \\
\hline
\endhead
\hline
\endfoot
\multicolumn{9}{|c|}{ORCA4 period} \\
\hline
GW190706\_222641 &   BBH &   72.2 \relax &  0.050 \relax &   0 \relax &   {-} \relax &  2.2e+02 \relax &  2.5e+58 \relax &  2.6e+03 \relax \\
GW190707\_093326 &   BBH &   70.7 \relax &  0.055 \relax &   0 \relax &   {-} \relax &  1.5e+02 \relax &  5.0e+56 \relax &  3.3e+02 \relax \\
GW190708\_232457 &   BBH &   53.4 \relax &  0.110 \relax &   0 \relax &   {-} \relax &  2.9e+02 \relax &  7.2e+56 \relax &  3.1e+02 \relax \\
GW190719\_215514 &   BBH &   44.6 \relax &  0.064 \relax &   0 \relax &   {-} \relax &  2.1e+02 \relax &  2.1e+58 \relax &  4.7e+03 \relax \\
GW190725\_174728 &   BBH &   45.5 \relax &  0.059 \relax &   0 \relax &   {-} \relax &  1.7e+02 \relax &  1.2e+57 \relax &  1.1e+03 \relax \\
GW190727\_060333 &   BBH &   82.6 \relax &  0.045 \relax &   0 \relax &   {-} \relax &  1.1e+02 \relax &  6.0e+57 \relax &  1.1e+03 \relax \\
GW190728\_064510 &   BBH &   77.0 \relax &  0.050 \relax &   0 \relax &   {-} \relax &  1.6e+02 \relax &  5.5e+56 \relax &  3.3e+02 \relax \\
GW190731\_140936 &   BBH &   77.7 \relax &  0.057 \relax &   0 \relax &   {-} \relax &  1.0e+02 \relax &  8.1e+57 \relax &  1.6e+03 \relax \\
GW190803\_022701 &   BBH &   48.8 \relax &  0.102 \relax &   0 \relax &   {-} \relax &  5.4e+02 \relax &  3.2e+58 \relax &  6.8e+03 \relax \\
GW190805\_211137 &   BBH &   72.7 \relax &  0.070 \relax &   0 \relax &   {-} \relax &  1.4e+02 \relax &  4.5e+58 \relax &  7.0e+03 \relax \\
GW190814         &  NSBH &  100.0 \relax &  0.038 \relax &   0 \relax &   {-} \relax &  1.3e+02 \relax &  3.0e+55 \relax &  6.8e+01 \relax \\
GW190828\_063405 &   BBH &   62.4 \relax &  0.057 \relax &   1 \relax &  0.06 \relax &  2.9e+02 \relax &  5.7e+57 \relax &  1.1e+03 \relax \\
GW190828\_065509 &   BBH &   77.1 \relax &  0.067 \relax &   0 \relax &   {-} \relax &  1.0e+02 \relax &  1.3e+57 \relax &  5.8e+02 \relax \\
GW190909\_114149 &   BBH &   52.5 \relax &  0.076 \relax &   0 \relax &   {-} \relax &  1.9e+02 \relax &  2.7e+58 \relax &  5.0e+03 \relax \\
GW190915\_235702 &   BBH &   90.2 \relax &  0.061 \relax &   0 \relax &   {-} \relax &  3.7e+02 \relax &  5.0e+57 \relax &  1.1e+03 \relax \\
GW190916\_200658 &   BBH &   56.1 \relax &  0.075 \relax &   0 \relax &   {-} \relax &  1.5e+02 \relax &  4.0e+58 \relax &  6.6e+03 \relax \\
GW190917\_114630 &  NSBH &   72.8 \relax &  0.066 \relax &   0 \relax &   {-} \relax &  1.8e+02 \relax &  4.9e+56 \relax &  1.4e+03 \relax \\
GW190924\_021846 &   BBH &   40.3 \relax &  0.043 \relax &   0 \relax &   {-} \relax &  1.6e+02 \relax &  2.6e+56 \relax &  2.7e+02 \relax \\
GW190925\_232845 &   BBH &  100.0 \relax &  0.059 \relax &   0 \relax &   {-} \relax &  1.4e+02 \relax &  5.9e+56 \relax &  1.9e+02 \relax \\
GW190926\_050336 &   BBH &   52.6 \relax &  0.052 \relax &   0 \relax &   {-} \relax &  1.6e+02 \relax &  2.9e+58 \relax &  6.2e+03 \relax \\
GW190929\_012149 &   BBH &   66.5 \relax &  0.075 \relax &   0 \relax &   {-} \relax &  1.6e+02 \relax &  1.3e+58 \relax &  2.0e+03 \relax \\
GW190930\_133541 &   BBH &   22.1 \relax &  0.073 \relax &   0 \relax &   {-} \relax &  4.6e+02 \relax &  1.4e+57 \relax &  9.9e+02 \relax \\
GW191103\_012549 &   BBH &   43.0 \relax &  0.063 \relax &   0 \relax &   {-} \relax &  4.0e+02 \relax &  2.4e+57 \relax &  1.6e+03 \relax \\
GW191105\_143521 &   BBH &   85.3 \relax &  0.067 \relax &   0 \relax &   {-} \relax &  1.5e+02 \relax &  9.5e+56 \relax &  6.9e+02 \relax \\
GW191109\_010717 &   BBH &   88.4 \relax &  0.070 \relax &   0 \relax &   {-} \relax &  1.3e+02 \relax &  1.5e+57 \relax &  2.3e+02 \relax \\
GW191113\_071753 &   BBH &   71.8 \relax &  0.070 \relax &   0 \relax &   {-} \relax &  1.3e+02 \relax &  1.7e+57 \relax &  1.3e+03 \relax \\
GW191127\_050227 &   BBH &   42.5 \relax &  0.079 \relax &   0 \relax &   {-} \relax &  1.4e+02 \relax &  1.5e+58 \relax &  2.4e+03 \relax \\
GW191129\_134029 &   BBH &   62.4 \relax &  0.059 \relax &   0 \relax &   {-} \relax &  1.2e+02 \relax &  3.2e+56 \relax &  2.5e+02 \relax \\
GW191204\_110529 &   BBH &   45.1 \relax &  0.084 \relax &   0 \relax &   {-} \relax &  1.4e+02 \relax &  4.9e+57 \relax &  1.6e+03 \relax \\
GW191204\_171526 &   BBH &   84.6 \relax &  0.041 \relax &   0 \relax &   {-} \relax &  1.2e+02 \relax &  2.1e+56 \relax &  1.3e+02 \relax \\
GW191215\_223052 &   BBH &   64.2 \relax &  0.052 \relax &   0 \relax &   {-} \relax &  1.6e+02 \relax &  3.1e+57 \relax &  9.8e+02 \relax \\
GW191219\_163120 &  NSBH &   70.0 \relax &  0.065 \relax &   0 \relax &   {-} \relax &  1.7e+02 \relax &  3.6e+56 \relax &  2.4e+03 \relax \\
GW191222\_033537 &   BBH &   74.5 \relax &  0.054 \relax &   0 \relax &   {-} \relax &  1.1e+02 \relax &  6.2e+57 \relax &  1.1e+03 \relax \\
GW191230\_180458 &   BBH &   58.6 \relax &  0.054 \relax &   0 \relax &   {-} \relax &  1.1e+02 \relax &  1.2e+58 \relax &  1.9e+03 \relax \\
GW200105\_162426 &  NSBH &   52.5 \relax &  0.158 \relax &   0 \relax &   {-} \relax &  1.9e+02 \relax &  6.7e+55 \relax &  1.7e+02 \relax \\
GW200112\_155838 &   BBH &   45.1 \relax &  0.116 \relax &   0 \relax &   {-} \relax &  1.9e+02 \relax &  1.1e+57 \relax &  2.1e+02 \relax \\
\hline
\multicolumn{9}{|c|}{ORCA6 period} \\
\hline
GW200128\_022011 &   BBH &   60.1 \relax &  0.255 \relax &   0 \relax &   {-} \relax &  6.8e+01 \relax &  4.8e+57 \relax &  8.7e+02 \relax \\
GW200129\_065458 &   BBH &    8.5 \relax &  0.175 \relax &   0 \relax &   {-} \relax &  7.5e+01 \relax &  3.0e+56 \relax &  5.7e+01 \relax \\
GW200208\_130117 &   BBH &  100.0 \relax &  0.127 \relax &   0 \relax &   {-} \relax &  5.3e+01 \relax &  1.3e+57 \relax &  2.6e+02 \relax \\
GW200209\_085452 &   BBH &   42.8 \relax &  0.358 \relax &   0 \relax &   {-} \relax &  1.2e+02 \relax &  8.6e+57 \relax &  1.9e+03 \relax \\
GW200210\_092254 &  NSBH &   27.2 \relax &  0.258 \relax &   0 \relax &   {-} \relax &  7.6e+01 \relax &  3.2e+56 \relax &  6.8e+02 \relax \\
GW200219\_094415 &   BBH &   90.5 \relax &  0.213 \relax &   0 \relax &   {-} \relax &  7.6e+01 \relax &  4.7e+57 \relax &  1.0e+03 \relax \\
GW200220\_061928 &   BBH &   59.7 \relax &  0.290 \relax &   0 \relax &   {-} \relax &  6.2e+01 \relax &  1.8e+58 \relax &  1.8e+03 \relax \\
GW200220\_124850 &   BBH &   53.5 \relax &  0.286 \relax &   0 \relax &   {-} \relax &  7.6e+01 \relax &  1.0e+58 \relax &  2.2e+03 \relax \\
GW200302\_015811 &   BBH &   60.6 \relax &  0.497 \relax &   1 \relax &  0.39 \relax &  2.1e+02 \relax &  2.9e+57 \relax &  7.2e+02 \relax \\
GW200306\_093714 &   BBH &   48.9 \relax &  0.367 \relax &   0 \relax &   {-} \relax &  2.1e+02 \relax &  7.7e+57 \relax &  2.5e+03 \relax \\
GW200308\_173609 &   BBH &   55.3 \relax &  0.335 \relax &   1 \relax &  0.28 \relax &  1.2e+02 \relax &  1.3e+59 \relax &  4.5e+04 \relax \\
GW200316\_215756 &   BBH &    3.2 \relax &  0.261 \relax &   1 \relax &  0.23 \relax &  7.9e+01 \relax &  5.4e+56 \relax &  3.7e+02 \relax \\
GW200322\_091133 &   BBH &   49.5 \relax &  0.376 \relax &   0 \relax &   {-} \relax &  1.7e+02 \relax &  1.3e+59 \relax &  3.7e+05 \relax \\
\end{longtable}
\end{center}

\subsection*{Individual limits for the 5--30 MeV energy range}

\begin{figure}[hbtp]
    \centering
    \includegraphics[width=.7\linewidth]{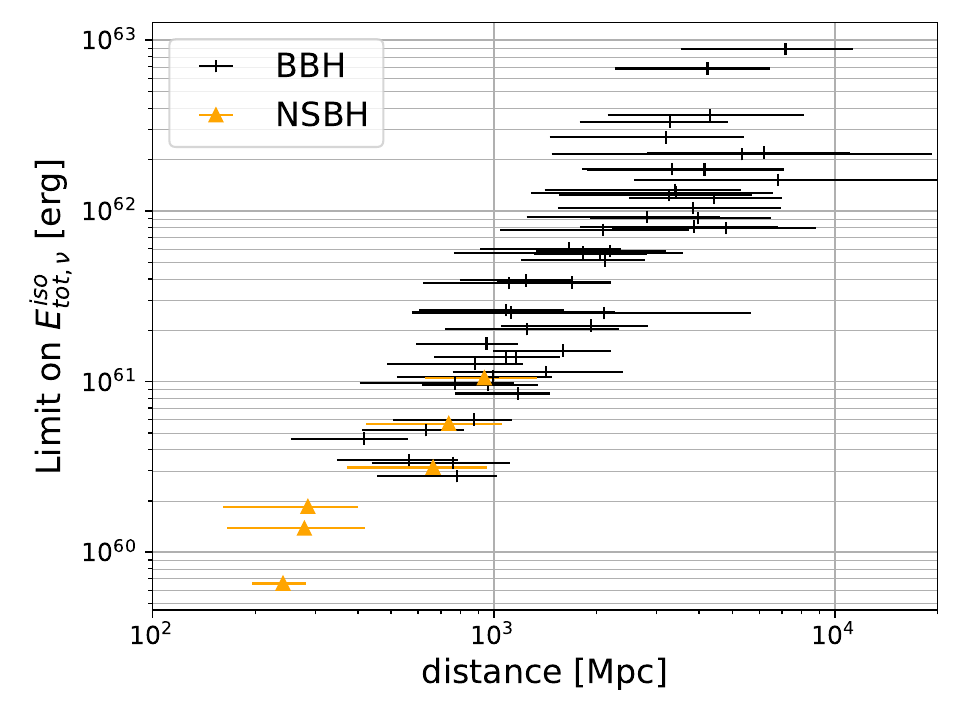}
    \caption{90\% upper limits on the total energy $E^{\rm iso}_{\rm tot, \nu}$ emitted in neutrinos assuming a quasi-thermal distribution centered at \qty{13}{\mega\electronvolt}, as a function of the source luminosity distance, assuming an $E^{-2}$ spectrum and isotropic emission.}
    \label{fig:limits_mev}
\end{figure}

Only four events have a p-value lower than $0.2$, with a minimum of $0.02$, which is fully compatible with the background expectation. For the considered \qty{2}{\second} time window, the upper limits on the neutrino fluence range between $\num{1e10}$ and \qty{1e11}{\per\square\centi\meter}, and on the total energy emitted in neutrinos between \num{e60} and \qty{e63}{\erg}, as reported in \autoref{fig:limits_mev}. Given that these limits are not very constraining with respect to the total available energy budget in the merger ($\lesssim$ \qtyrange{e55}{e56}{\erg}), stacking limits have not been estimated for this energy range.

\subsection*{Individual limits for the GeV--TeV energy range} 

\begin{figure}[hbtp]
    \centering
    \includegraphics[width=0.7\textwidth]{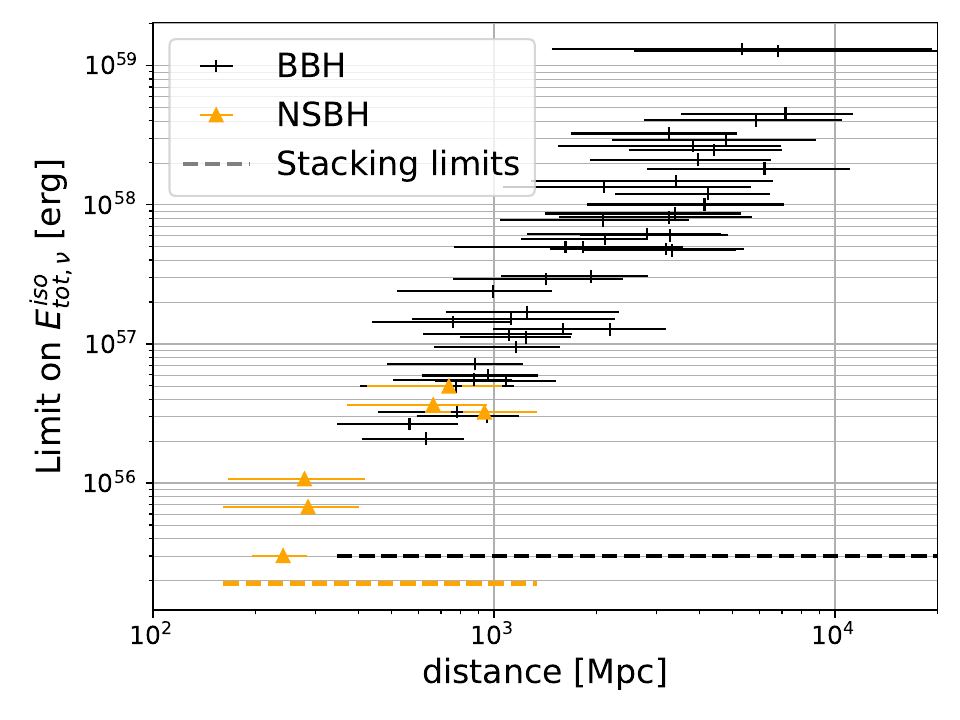}
    \includegraphics[width=0.7\textwidth]{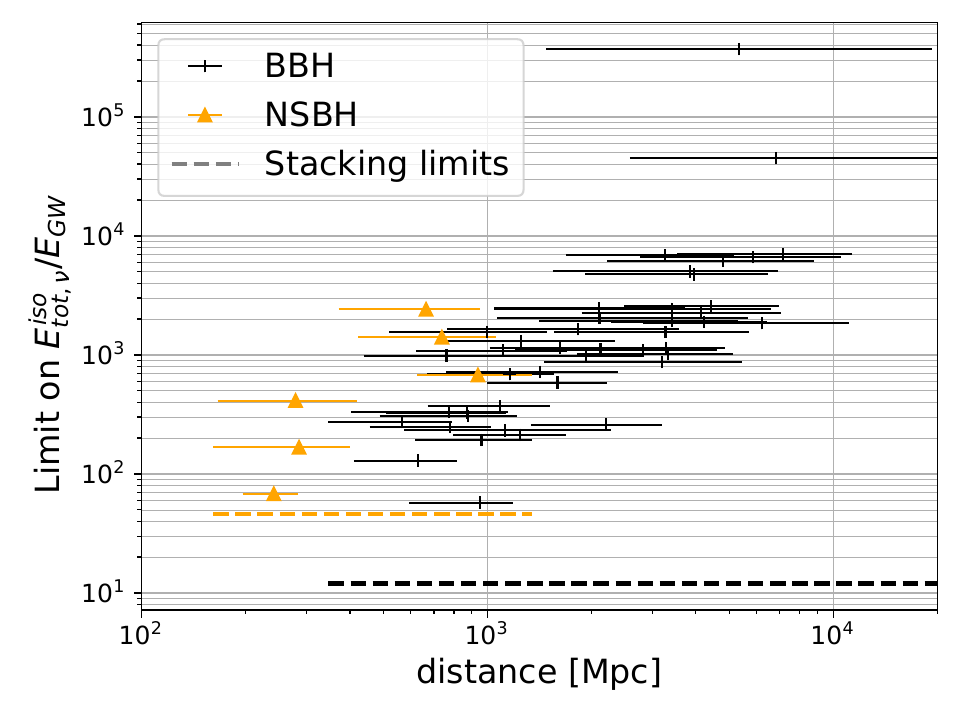}
    \caption{90\% upper limits on the total energy $E^{\rm iso}_{\rm tot, \nu}$ emitted in neutrinos of all flavors (top) and on the ratio $f^{\rm iso}_\nu = E^{\rm iso}_{\rm tot, \nu} / E_{\rm GW}$ (bottom) as a function of the source luminosity distance, assuming an $E^{-2}$ spectrum and isotropic emission. The horizontal bars indicate the 5-95\% range of the luminosity distance estimate, and the markers/colors correspond to the different source categories. The dashed bars correspond to the upper limits from the stacking analysis for BBH and NSBH categories.}
    \label{fig:results:HE:Eiso}
\end{figure}

The obtained upper limits on the total all-flavor isotropic energy $E^{\rm iso}_{\rm tot, \nu}$ and on the ratio between the energy radiated in neutrinos and in GWs $f^{\rm iso}_\nu = E^{\rm iso}_{\rm tot, \nu} / E_{\rm GW}$, for each of the 44 BBH and the 6 NSBH events, are shown in \autoref{fig:results:HE:Eiso}.

The most constraining individual limits for the BBH events are $E_{\rm tot,\nu}^{{\rm iso,} 90\%} = \qty{2.1e56}{\erg}$ (for GW191204\_171526) and $f_\nu^{{\rm iso,} 90\%} = 57$ (for GW200129\_065458). Similarly, the most constraining individual NSBH event is the closest one, GW190814, with limits $E_{\rm tot,\nu}^{{\rm iso,} 90\%} = \qty{3.0e55}{\erg}$ and $f_\nu^{{\rm iso,} 90\%} = 68$.

\subsection*{Limits from the stacking analysis in the GeV--TeV range}

A stacking analysis is performed using the 44 BBH events followed up with KM3NeT. Assuming that all BBH emit the same typical total energy in neutrinos, a limit of $E_{\rm tot,\nu}^{{\rm iso,} 90\%} = \qty{3.0e55}{\erg}$ is obtained. Assuming instead that all sources have the same ratio between neutrino and GW emission, the limit is $f_\nu^{{\rm iso,} 90\%} = 12$. The stacked total energy limit is seven times better than the most constraining individual event in this category.

The stacking of the six NSBH candidates followed with KM3NeT gives $E_{\rm tot,\nu}^{{\rm iso,} 90\%} = \qty{1.9e55}{\erg}$ and $f_\nu^{{\rm iso,} 90\%} = 46$. Given the small number of NSBH events in the catalogs, the gain relative to the individual limits is moderate.

\section{Discussion and conclusions}
\label{sec:concl}

\begin{figure}[hbtp]
    \centering
    \includegraphics[width=0.7\linewidth]{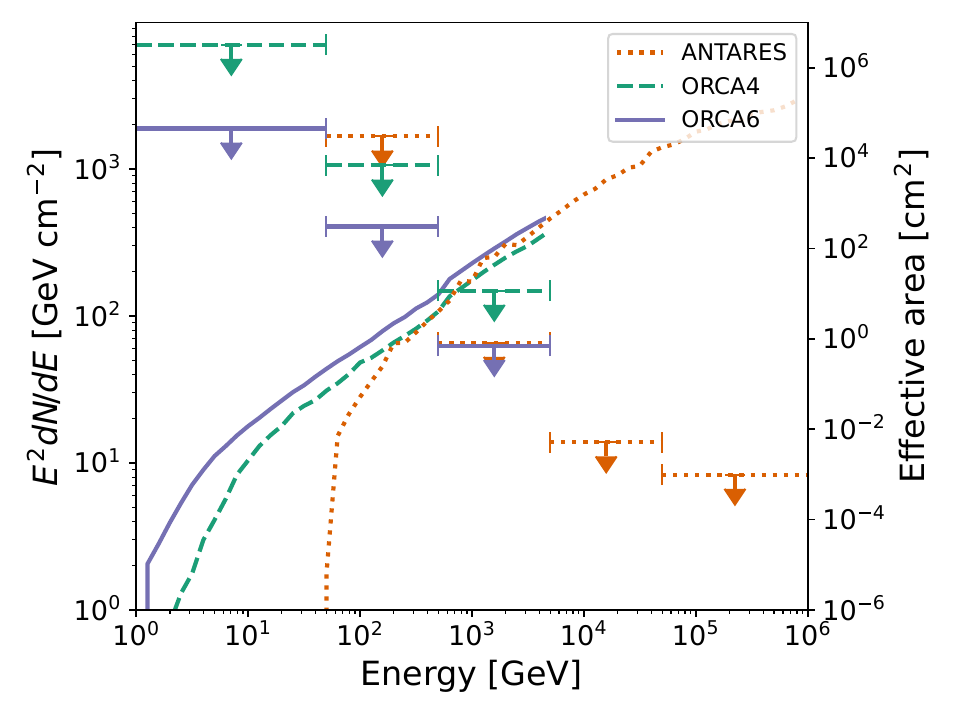}
    \caption{Comparison of ORCA effective areas at upgoing track selection level with ANTARES~\cite{ANTARES:2023wcj}, for GW190814 (for ANTARES and ORCA4) and GW200208\_130117 (for ORCA6). The two GW events have very similar sky coverage and thus comparable limits. The differential upper limits (horizontal lines) were obtained by considering independently bins in true neutrino energy and computing the corresponding limit on the flux normalization assuming an $E^{-2}$ spectrum only within each bin (and zero elsewhere).}
    \label{fig:comp_aeff}
\end{figure}

\begin{figure}[hbtp]
    \centering
    \includegraphics[width=0.7\linewidth]{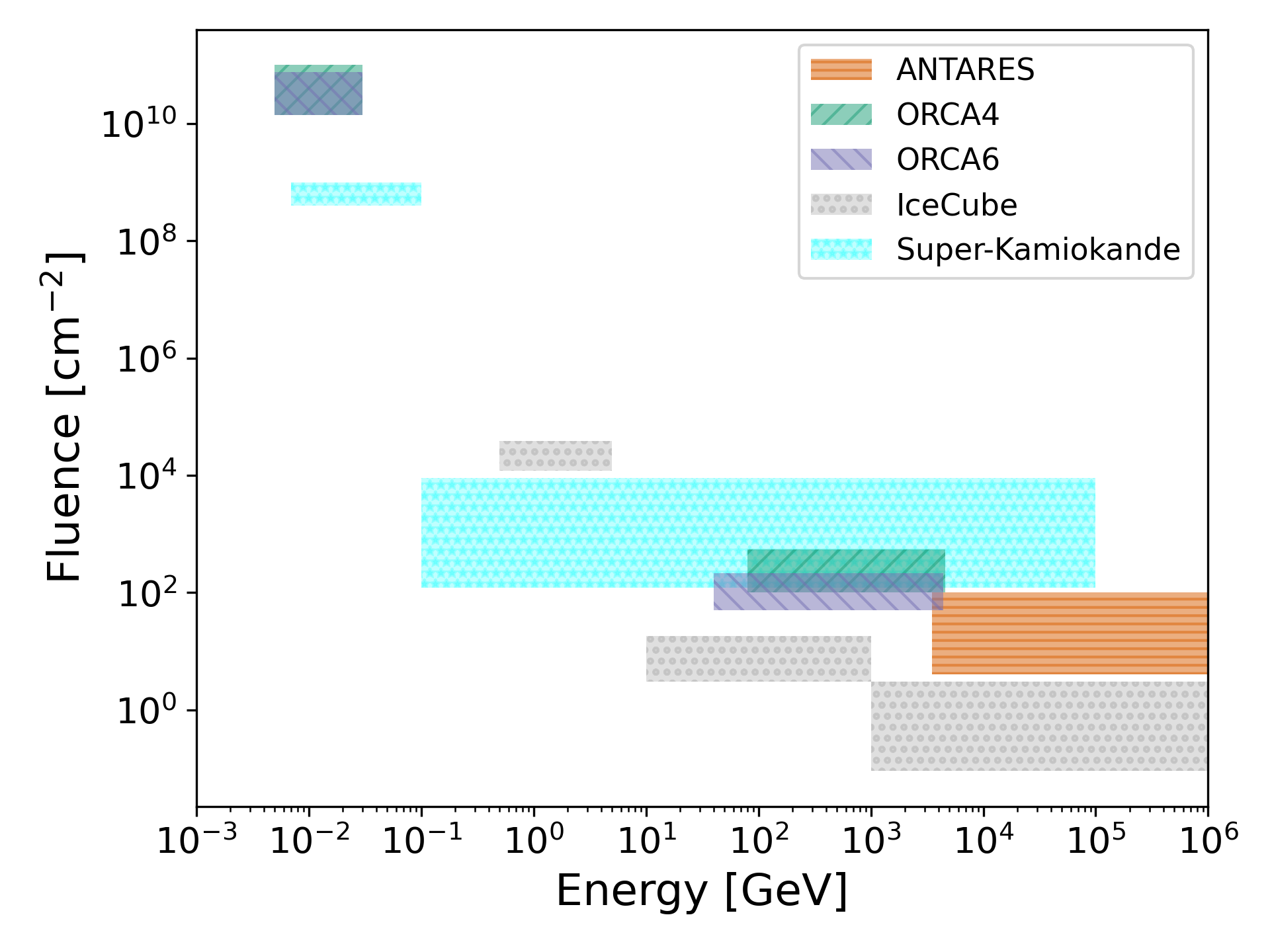}
    \caption{Range of 90\% upper limits on the total neutrino fluence for both analyses. For MeV-scale neutrinos, these are directly the limits reported in \autoref{tab:results:VLE}. For all reported results above $\qty{1}{\giga\electronvolt}$, the fluence is computed integrating energies above \qty{1}{\giga\electronvolt} ($\Phi = \int_{\qty{1}{\giga\electronvolt}} \phi E^{-2} {\rm d}E$), and the horizontal widths of the bands delimit the central energy range expected to contain 90\% of the signal events (except for IceCube and Super-Kamiokande results where the full sensitive range is shown). The ANTARES limits are reported in \cite{ANTARES:2023wcj}. The IceCube results are extracted from \cite{IceCube:2021ddq}, \cite{IceCube:2021ahz}, and \cite{IceCube:2022mma}, from left to right. The Super-Kamiokande results are obtained from \cite{Super-Kamiokande:2021dav,Super-Kamiokande:2021_4724823}.}
    \label{fig:comp_limits}
\end{figure}

The limits presented in this paper cover two neutrino energy ranges: \qtyrange{5}{30}{\mega\electronvolt} and \unit{\giga\electronvolt}-\unit{\tera\electronvolt}. \autoref{fig:comp_aeff} illustrates the typical differential limits as a function of neutrino energy for ORCA4, ORCA6, and ANTARES~\cite{ANTARES:2023wcj}, overlaid with the corresponding detector effective areas. For the two ORCA configurations, the bin at the highest neutrino energies is always the most constraining as the expected number of signal events in this bin is still relatively high (as shown in the top right panel of \autoref{fig:selec_perfs}). The constraints derived here, with only a very partial ORCA detector (3-5\% of the total number of lines to be deployed), already bring complementary information as the two ORCA configurations are sensitive in a lower energy range than ANTARES and they provide better differential limits in that region of the spectrum. The small size of the ORCA4 and ORCA6 configurations, combined with this difference in terms of energy range, lead to worse integrated limits when comparing to ANTARES or to IceCube high-energy limits \cite{IceCube:2022mma}, as illustrated on \autoref{fig:comp_limits}. 

In the \unit{\mega\electronvolt} range, the obtained limits are of the same order as the ones reported by KamLAND \cite{KamLAND:2020ses}, although one to two orders of magnitude worse than Super-Kamiokande's \cite{Super-Kamiokande:2021dav}, as shown on \autoref{fig:comp_limits}.

As of autumn 2023, 18 lines are operating for ORCA and 28 for ARCA, with more lines scheduled to be deployed later in 2023 and in the following years. During the following GW observation campaigns, especially O4 which has started in spring 2023, follow-ups will be performed with much larger detectors than discussed in this article, leading to improved sensitivities and an extended energy range coverage. More detailed neutrino emission models may also be explored, beyond the isotropic $E^{-2}$ and quasi-thermal spectra investigated in the present study.

The ARCA configuration, which did not contribute to the present results for O3, is expected to participate for the first time in the follow-ups during O4. Its energy coverage at very high energy ($\gtrsim \unit{\tera\electronvolt}$) is complementary to ORCA, hence enhancing KM3NeT sensitivity and the discovery lever arm, especially for hard spectra. As the field of view of KM3NeT is very different from that of IceCube, even partial KM3NeT detectors will be able to contribute significantly to the searches, especially for sources localized in the Southern Sky.

For \unit{\mega\electronvolt} neutrinos, the gain is directly proportional to the size of the detector, as outlined in \autoref{eq:limits:vle}, and KM3NeT is expected to reach similar sensitivities as Super-Kamiokande by the end of the construction.

The KM3NeT telescope is also performing real-time follow-ups during O4, planning to release results as fast as possible to help constrain the localization of a potential joint source and guide electromagnetic observations. It will improve the chance of identifying the corresponding electromagnetic emission and thus eventually constrain source models, jet structure, and production mechanisms.

Furthermore, the increasing number of detected GW sources, especially binaries involving neutron stars, will enhance the capability of stacking analyses. Even in the absence of individually significant sources, some hints of neutrino emission may arise for a sub-population of these sources, as a slight deviation from background-only predictions. Though the underlying production mechanisms are very different, covering different energy ranges from \unit{\mega\electronvolt} to \unit{\peta\electronvolt} with KM3NeT may help reveal the nature of the sources or identify sub-populations.

\acknowledgments{The authors acknowledge the financial support of the funding agencies:
Agence Nationale de la Recherche (contract ANR-15-CE31-0020), Centre National de la Recherche Scientifique (CNRS), Commission Europ\'eenne (FEDER fund and Marie Curie Program), LabEx UnivEarthS (ANR-10-LABX-0023 and ANR-18-IDEX-0001), Paris \^Ile-de-France Region, France;
Shota Rustaveli National Science Foundation of Georgia (SRNSFG, FR-22-13708), Georgia;
The General Secretariat of Research and Innovation (GSRI), Greece;
Istituto Nazionale di Fisica Nucleare (INFN), Ministero dell'Universit\`a e della Ricerca (MIUR), PRIN 2017 program (Grant NAT-NET 2017W4HA7S) Italy;
Ministry of Higher Education, Scientific Research and Innovation, Morocco, and the Arab Fund for Economic and Social Development, Kuwait;
Nederlandse organisatie voor Wetenschappelijk Onderzoek (NWO), the Netherlands;
The National Science Centre, Poland (2021/41/N/ST2/01177); The grant “AstroCeNT: Particle Astrophysics Science and Technology Centre”, carried out within the International Research Agendas programme of the Foundation for Polish Science financed by the European Union under the European Regional Development Fund;
National Authority for Scientific Research (ANCS), Romania;
Grants PID2021-124591NB-C41, -C42, -C43 funded by MCIN/AEI/ 10.13039/501100011033 and, as appropriate, by “ERDF A way of making Europe”, by the “European Union” or by the “European Union NextGenerationEU/PRTR”, Programa de Planes Complementarios I+D+I (refs. ASFAE/2022/023, ASFAE/2022/014), Programa Prometeo (PROMETEO/2020/019) and GenT (refs. CIDEGENT/2018/034, /2019/043, /2020/049. /2021/23) of the Generalitat Valenciana, Junta de Andaluc\'{i}a (ref. SOMM17/6104/UGR, P18-FR-5057), EU: MSC program (ref. 101025085), Programa Mar\'{i}a Zambrano (Spanish Ministry of Universities, funded by the European Union, NextGenerationEU), Spain;
The European Union's Horizon 2020 Research and Innovation Programme (ChETEC-INFRA - Project no. 101008324); Fonds de la Recherche Scientifique - FNRS, Belgium; Francqui foundation.}

\bibliographystyle{JHEP}
\bibliography{references}

\end{document}